\PassOptionsToPackage{pdfpagelabels=false}{hyperref}   
\expandafter\def\csname ver@fixltx2e.sty\endcsname{}   

\documentclass[fleqn,usenatbib]{mnras}

\usepackage{graphicx}	
\usepackage{amsmath}	
\usepackage{amssymb}	
\usepackage{siunitx}
\usepackage{makecell}

\DeclareSIUnit\angstrom{\text{Å}}

\usepackage{txfonts}

\usepackage[T1]{fontenc}
\usepackage{ae,aecompl}

\title[Chemical co-evolution of gas and stars]{SDSS-IV MaNGA: the chemical co-evolution of gas and stars in spiral galaxies}

\author[M. J. Greener et al.]{
Michael J. Greener,$^{1}$\thanks{E-mail: michael.greener@nottingham.ac.uk}
Alfonso Arag{\'o}n-Salamanca,$^{1}$
Michael Merrifield,$^{1}$
Thomas Peterken,$^{1}$
\newauthor
Elizaveta Sazonova,$^{2}$
Roan Haggar,$^{1}$
Dmitry Bizyaev,$^{3, \: 4}$
Joel R. Brownstein,$^{5}$
Richard R. Lane,$^{6}$
\newauthor
and Kaike Pan$^{3}$
\\
$^{1}$School of Physics \& Astronomy, University of Nottingham, University Park, Nottingham, NG7 2RD, UK\\
$^{2}$Department of Physics and Astronomy, Johns Hopkins University, Baltimore, MD 21218, USA\\
$^{3}$Apache Point Observatory and New Mexico State University, P.O. Box 59, Sunspot, NM 88349, USA\\
$^{4}$Sternberg Astronomical Institute, Moscow State University, Universitetskiy Prospekt 13, Moscow, 119992, Russia\\
$^{5}$Department of Physics and Astronomy, University of Utah, Salt Lake City, UT 84112, USA\\
$^{6}$Centro de Investigaci{\'o}n en Astronom{\'i}a, Universidad Bernardo O'Higgins, Avenida Viel 1497, Santiago, Chile\\
}

\date{Accepted XXX. Received YYY; in original form ZZZ}

\pubyear{2022}

\begin{document}
\label{firstpage}
\pagerange{\pageref{firstpage}--\pageref{lastpage}}
\maketitle

\begin{abstract}
We investigate archaeologically how the metallicity in both stellar and gaseous components of spiral galaxies of differing masses evolve with time, using data from the SDSS-IV MaNGA survey. For the stellar component, we can measure this evolution directly by decomposing the galaxy absorption-line spectra into populations of different ages and determining their metallicities. For the gaseous component, we can only measure the present-day metallicity directly from emission lines. However, there is a well-established relationship between gas metallicity, stellar mass and star formation rate which does not evolve significantly with redshift; since the latter two quantities can be determined directly for any epoch from the decomposition of the absorption-line spectra, we can use this relationship to infer the variation in gas metallicity over cosmic time. Comparison of present-day values derived in this way with those obtained directly from the emission lines confirms the validity of the method. Application of this approach to a sample of 1619 spiral galaxies reveals how the metallicity of these systems has changed over the last 10 billion years since cosmic noon. For lower-mass galaxies, both stellar and gaseous metallicity increase together, as one might expect in well-mixed fairly isolated systems. In higher-mass systems, the average stellar metallicity has not increased in step with the inferred gas metallicity, and actually decreases with time. Such disjoint behaviour is what one might expect if these more massive systems have accreted significant amounts of largely pristine gas over their lifetimes, and this material has not been well mixed into the galaxies.
\end{abstract}




\section{Introduction}
\label{sec:Introduction}

Measuring the levels of heavy elements within galaxies over time is fundamental in order to fully understand their evolution. Typically, the chemical composition of galaxies is quantified by determining the levels of elements heavier than helium -- the ``metallicities'' -- of the stars and gas\footnote{While stellar metallicities can readily be determined via spectral fitting methods, gas metallicities are not directly measured. Instead, we calculate the relative abundance of oxygen to hydrogen of the gas, defined in units of $12 + \log(\rm O / H)$. Scaling relations are then used to obtain estimates of galactic gas metallicities from these abundances; we describe this process in detail in Section~\ref{subsubsec:Emission_lines}.} within these galaxies. Recent reviews by \citet{Kewley2019UnderstandingLines} and \citet{Maiolino2019DeGalaxies} comprehensively discuss galactic gas metallicities, and \citet{Madau2014CosmicHistory} and \citet{Maiolino2019DeGalaxies} review metallicities of the stellar populations within galaxies. While most studies of chemical evolution tend to focus on gas and stellar metallicities separately, these quantities can -- and should -- be treated on equal footing, since the stars whose metallicities we measure form from the same gas we are also interested in studying. Previous authors, such as \citet{Lian2018TheIMF, Lian2018Modelling0}, and \citet{Yates2021L-GALAXIESGalaxies}, have found success in modelling the metallicities of the gas and the stars within galaxies in tandem. The results from this paper will aid the efforts of such authors by providing observational evidence to feed into these models.

Metallicities are known to evolve with cosmic time. However, the precise evolution of the metal content of both the gas and the stellar populations within galaxies is still not fully understood. Since metallicities are closely associated with other fundamental galaxy properties such as star formation rates (SFRs) and stellar masses \citep[e.g.][]{Mannucci2010AGalaxies, Yates2012TheData, Curti2020TheGalaxies}, this is a crucial problem for astronomy to solve. Consequently, the evolution of metallicities back to the point at which galactic SFRs were at their most prodigious is of particular interest. This epoch, known as ``cosmic noon'', occurred at a redshift of $z \sim 1.5$ \citep{Madau1998TheGalaxies, Madau2014CosmicHistory}. It is thus especially important to investigate how metallicities have evolved since cosmic noon to build a robust model of galaxy evolution.

\medskip

In recent years, stellar metallicity histories have been determined thanks to observations of low-redshift, spatially resolved galaxies made by integral field unit (IFU) surveys such as the Mapping Nearby Galaxies at Apache Point Observatory \citep[MaNGA;][]{Bundy2015OVERVIEWOBSERVATORY} survey; the Calar Alto Legacy Integral Field Area \citep[CALIFA;][]{Sanchez2012CALIFASurvey} survey; and the Sydney-AAO (Australian Astronomical Observatory) Multi-object Integral field spectrograph \citep[SAMI;][]{Croom2012TheSpectrograph} galaxy survey. For example, \citet{Peterken2019Time-slicingMaNGA, Peterken2020SDSS-IVGalaxies}, \citet{Camps-Farina2021EvolutionGalaxies, Camps-Farina2022ChemicalGalaxies}, and \citet{Fraser-McKelvie2022TheGalaxies} estimate spatially resolved stellar metallicity histories by excavating the fossil records of nearby MaNGA, CALIFA, and SAMI galaxies respectively. Spectral fitting methods, such as those employed by these authors, allow us to track the evolution of the average stellar metallicity with time within individual galaxies. This evolution can be yet further constrained when such methods are used in conjunction with direct observations of average stellar metallicities in higher-redshift galaxies (see, for instance, \citealp{Onodera2015THE1.6, Sanders2021The3.3}; and \citealp{Beverage2021ElementalQuenching}).

By contrast, determining the gas metallicity histories within galaxies has previously proved elusive. The gas metallicity of a given galaxy can be estimated at the time at which the light was emitted by measuring certain emission lines in the galactic spectra, but it is not possible to track the evolution of these emission lines back over cosmic time. Nevertheless, the desire to do so is well-motivated, as work by \citet{ValeAsari2007TheSurvey} demonstrates. From their analysis of the star formation histories (SFHs) of over 80~000 galaxies, these authors found that galaxies with lower mean gas metallicities evolved more slowly than metal-rich galaxies.

It is, of course, possible to model the gas metallicity histories of galaxies through the use of simulations, as work by authors such as \citet{Fu2012TheGalaxies}, \citet{Somerville2015StarGas} and \citet{Yates2021L-GALAXIESGalaxies} demonstrates. Observationally, however, the evolution of the gas metallicity has traditionally been estimated by observing galaxies at increasing redshift (see, for example, work by \citealp{Troncoso2014Metallicity3.4, Wuyts2014A2.6}; and \citealp{Kashino2017THEMEDIUM}). Unfortunately, this approach provides us only with snapshots of the chemical evolution of galaxies; furthermore, the quality of such snapshots diminishes with the redshift at which they are obtained. In theory, it should be possible to track the complete evolution histories of the gas metallicity in low-redshift, spatially resolved galaxies, just as is done for the average stellar metallicity. However, the wealth of observational information from IFU surveys such as MaNGA has not yet been fully exploited to determine the fossil records of the gas metallicities, as has been so successfully done for the stars which formed from this same gas.

Fortunately, gas metallicities are closely linked to other physical properties of galaxies. In particular, star-forming galaxies with higher stellar masses exhibit higher SFRs \citep[e.g.][]{Brinchmann2004TheUniverse, Noeske2007StarGalaxies, Whitaker2012THE2.5, Greener2020SDSS-IVGalaxies}. This correlation is often called the ``star formation main sequence'' (SFMS). More massive galaxies also tend to have higher gas metallicities -- a correlation often referred to as the ``mass--gas metallicity relation'' ($\rm MZ_{g}R$; e.g. \citealp{Tremonti2004TheSDSS, Mannucci2010AGalaxies, Zahid2013TheGalaxies, Maiolino2019DeGalaxies}). Moreover, and perhaps unsurprisingly, authors such as \citet{Mannucci2010AGalaxies}, \citet{Lara-Lopez2010AGalaxies}, \citet{Yates2012TheData}, and \citet{Curti2020TheGalaxies} have demonstrated that there exists a more fundamental relation between stellar mass, gas metallicity, and SFR. In essence, these authors find that gas metallicity exhibits a very tight relationship with both stellar mass and SFR. This relationship is generally referred to as the fundamental metallicity relation \citep[FMR; see, for instance, Figure 2 of][]{Mannucci2010AGalaxies}. Since the FMR does not evolve with redshift out to at least $z \sim 2.5$ \citep{Mannucci2010AGalaxies, Sanders2021The3.3}, it is therefore possible to infer the gas metallicity at earlier cosmic times from quantities that are derivable from spectral fitting -- namely, stellar mass and SFR as a function of redshift.

It should be noted that the FMR is under debate, with certain authors questioning the validity of such a relation (see, for instance, \citealp{Kashino2016HIDE-AND-SEEKRELATION, Telford2016EXPLORINGRATE}; and \citealp{Cresci2019FundamentalGalaxies}). In this paper, we assume the existence of such an FMR, in line with various of its proponents, including \citet{Mannucci2010AGalaxies}, \citet{Lara-Lopez2010AGalaxies}, \citet{Yates2012TheData}, \citet{Sanders2018The2.3, Sanders2021The3.3}, and \citet{Curti2020TheGalaxies}. An extensive review of the FMR -- and its origin -- is given by \citet{Maiolino2019DeGalaxies}.

This process of spectral fitting (i.e. decomposing the integrated galaxy spectrum into stellar population spectra of single ages and stellar metallicities) brings us into dangerous territory, and due care must be taken during the procedure. We have to ensure that the spectra are reliably decomposed into the full two-dimensional stellar metallicity--age plane of the single stellar population (SSP) templates. We must perform checks to ensure that this spectral decomposition can be done safely, after which we can confidently proceed with our analysis of the stellar masses and SFHs produced as a result of the procedure.

\medskip

In this paper, we infer via galactic archaeology the evolution of both the gas and stellar metallicity over cosmic time for a well-defined sample of galaxies in the present-day Universe that have been observed by the MaNGA survey. These quantities are calculated for the same galaxies both at the present day and at a redshift of $z \sim 1.4$. The gas metallicity histories of the galaxies are determined by using the FMR of \citet{Mannucci2010AGalaxies}. Giving equal consideration to both the gas and the stars within the galaxies allows us to probe the redshift evolution of their gas and stellar metallicities simultaneously back to the epoch of cosmic noon.

The paper is structured as follows. In Section~\ref{sec:Observations} we outline technical details of the MaNGA survey and our sample selection. We explain how stellar and gas metallicities are determined -- and also describe the various checks and tests we perform on the SSP templates employed in this work -- in Section~\ref{sec:Metallicities}. The results of this paper are presented and discussed in Section~\ref{sec:Results}, and finally Section~\ref{sec:Conclusions} summarises our conclusions. In this work, we employ a \citet{Chabrier2003GalacticFunction} initial mass function (IMF), and we assume a $\rm \Lambda CDM$ cosmology with $\Omega_M = 0.3$, $\Omega_{\Lambda} = 0.7$, and $H_0 = 70 \: \rm km \ s^{-1} \ Mpc^{-1}$.

\section{Data and Analysis}
\label{sec:Observations}

\subsection{MaNGA}
\label{subsec:MaNGA}

MaNGA \citep{Bundy2015OVERVIEWOBSERVATORY} is part of the fourth generation of the Sloan Digital Sky Survey \citep[SDSS-IV;][]{Blanton2017SloanUniverse}, and has now completed spectroscopic observations for over 10~000 nearby galaxies \citep{Yan2016SDSS-IVQuality, Wake2017TheConsiderations}. Hexagonal IFU fibre bundles \citep{Law2015OBSERVINGSURVEY} are connected to a spectrograph \citep{Smee2013THESURVEY, Drory2015THETELESCOPE}, which in turn is mounted on the $2.5\,{\rm m}$ telescope at Apache Point Observatory \citep{Gunn2006TheSurvey}. MaNGA acquires spectra for each galaxy to a distance of at least 1.5 effective radii, meaning that most of the light from the observed galaxies is captured. The wavelength range of MaNGA is $3600 - 10300~\si{\angstrom}$, and it has a spectral resolution of $R \sim 2000$ \citep{Bundy2015OVERVIEWOBSERVATORY}. The Data Reduction Pipeline \citep[DRP;][]{Law2016TheSurvey} reduces and calibrates the raw data observed by MaNGA \citep{Yan2016SDSS-IV/MaNGA:TECHNIQUE}, and the Data Analysis Pipeline \citep[DAP;][]{Westfall2019TheOverview, Belfiore2019TheModeling} subsequently further processes these data to produce maps detailing the physical properties of each galaxy.

\subsection{Sample Selection}
\label{subsec:Sample}

In order to select a representative sample of spiral galaxies, we make use of classifications provided by the citizen science project Galaxy Zoo 2 \citep[GZ2;][]{Willett2013GalaxySurvey}. The process that we follow is essentially identical to that described in \citet{Peterken2020SDSS-IVGalaxies} and \citet{Greener2021SDSS-IVRevisited}, except that we exploit data from the most recent eleventh (and final) MaNGA Product Launch (MPL-11). For a more detailed discussion of the method adopted here, we refer the reader to \citet{Willett2013GalaxySurvey} and \citet{Hart2016GalaxyBias}.

GZ2 classifications are available for a total of 9315 MPL-11 galaxies. The first step is to remove from this sample 81 galaxies which are obscured by a star or other artifact. To ensure we are only selecting spiral galaxies, we choose only those for which $> 43\%$ of $N \geq 20$ respondents observed either spiral features or a disk in the galaxy (as recommended by \citealp{Willett2013GalaxySurvey}). This cut reduces the sample to 6727 candidates who may be spiral galaxies. Since we wish our sample to be comprised of unequivocally spiral galaxies, we also want to select only those that are oriented reasonably face-on. In addition to requiring that $> 80\%$ of $N \geq 20$ respondents determine that each galaxy is not edge-on \citep{Willett2013GalaxySurvey}, we also select only those galaxies which have a photometric axis ratio of $\frac{b}{a} \geq 0.5$ (corresponding to an inclination of $i \geq \ang{60}$). This constraint is somewhat stricter than that proposed by \citet{Hart2017GalaxyAngles}, and further reduces the sample size to reasonably face-on 2081 spiral galaxies. Finally, we reject any galaxies that were flagged for poor data quality by the DRP or had for any reason failed to produce the necessary DAP data sets. This leaves a final sample comprised of 1619 reasonably face-on spiral galaxies that are fit for analysis.

\section{Metallicities}
\label{sec:Metallicities}

The chemical compositions of both the stellar populations as well as the gaseous interstellar medium (ISM) within the sample galaxies are considered in this paper. In this section, we detail the processes by which the metallicities of both the stars and the gas are measured.

\subsection{Stellar Metallicities}
\label{subsec:Stellar}

The average stellar metallicities of the sample galaxies are calculated using the full-spectrum stellar population fitting code \texttt{STARLIGHT} \citep{CidFernandes2005Semi-empiricalMethod}. The \texttt{STARLIGHT} fitting process is very similar to that implemented by \citet{Greener2020SDSS-IVGalaxies, Greener2021SDSS-IVRevisited}, and is fully detailed by \citet[][including their Appendix~A]{Peterken2020SDSS-IVGalaxies}. This fitting process used in this work has already been extensively tested by these authors. The main steps relevant to this work are described below.

Each of the MaNGA spectra are fitted using a linear combination of SSP templates from the E-MILES library of \citet{Vazdekis2016UV-extendedGalaxies}, which in turn is based on the earlier MILES library of \citet{Vazdekis2010EvolutionarySystem}. No binning of neighbouring spaxels is done, since we wish to fully retain all spatial information. During the fitting procedure, we adopt a \citet{Chabrier2003GalacticFunction} IMF, and the ``Padova'' isochrones of \citet{Girardi2000Evolutionary0.03}. The E-MILES templates constitute nine ages $(\log(\rm age / yr) = 7.85, \: 8.15, \: 8.45, \: 8.75, \: 9.05, \: 9.35, \: 9.65, \: 9.95, \: 10.25)$ and six stellar metallicities $([\rm M / H]_{*} = -1.71, \: -1.31, \: -0.71, \: -0.40, \: +0.00, \allowbreak \: +0.22)$. To reproduce younger stellar populations, we include an additional six ages $(\log(\rm age / yr) = 6.8, \: 6.9, \: 7.0, \allowbreak \: 7.2, \: 7.4, \: 7.6)$ and two stellar metallicities $([\rm M / H]_{*} = -0.41, \: +0.00)$ from the templates of \citet{Asad2017YoungCMDs}\footnote{These SSP ages, which are frequently referred to as lookback times throughout this paper, are considered in the rest frame of a given galaxy. For reference, the median redshift of the sample is $z = 0.036$, which corresponds to a lookback time of ${\sim} 10^{8.7} \: \rm yr$.}. These younger templates are produced in the same way as the E-MILES templates, except the (very similar) isochrones of \citet{Bertelli1994TheoreticalOpacities.} are used. Following the advice of \citet{Ge2018RecoveringUncertainties} and \citet{CidFernandes2018OnAlgorithms}, we run \texttt{STARLIGHT} in a configuration which prioritises robustness over computation times.

As \citet{Peterken2020SDSS-IVGalaxies} recommend, we do not use stellar metallicities at ages younger than $10^{7.6} \: \rm yr$ for any subsequent analysis, due to uncertainties in the derived SFHs of such populations \citep[see also][]{CidFernandes2010TestingMethodology}. These uncertainties are likely related to the ``UV upturn'' produced by planetary nebulae in old stellar populations (see \citealp{Yi2008TheUpturn} for a review of this phenomenon). Furthermore, at ages younger than $10^{7.6} \: \rm yr$, the calculated stellar metallicities also become increasingly unreliable; young, metal-poor stars are rare, and thus it is difficult to create accurate SSP templates for such populations. Moreover, the optical spectra of very hot type O stars do not have significant metal absorption lines, and thus provide poor stellar metallicity diagnostics. We do, however, include in our analysis the SSP templates at this threshold age of $10^{7.6} \: \rm yr$. We do not use stellar metallicities in the age bin at $10^{10.25} \: \rm yr$, since this is older than the age of the Universe. As such, it makes no physical sense to retain this age bin for our analysis.

The \texttt{STARLIGHT} fitting process generates a set of weights which encode the mass contribution made by each SSP to the light seen in that spectrum for every spaxel across the face of each galaxy. Summing the results from each spaxel then gives a fit to the integrated light from the entire galaxy, reducing the data from each galaxy to a two-dimensional parameter space of mass-weighted stellar metallicity and age. The contributions from SSPs of different mass-weighted stellar metallicities are then summed. This step further collapses the data for each galaxy to a one-dimensional function that encodes the proportion of stellar mass that stars in different age bins contribute to the total mass of that galaxy.

\begin{figure}
	\includegraphics[width=0.47\textwidth]{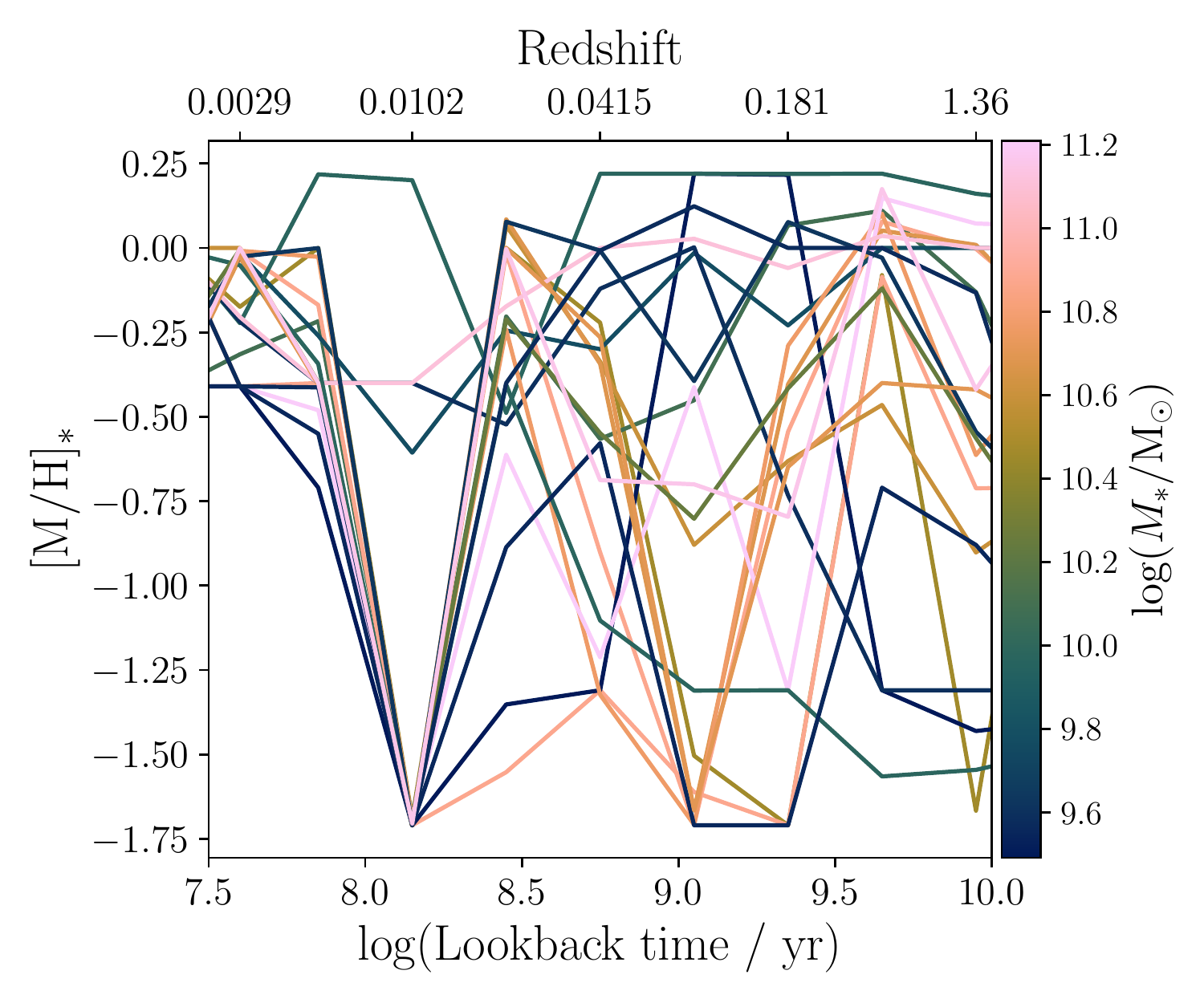}
    \caption{Mass-weighted mean stellar metallicity $[\rm M / H]_{*}$ in each age bin, determined by \texttt{STARLIGHT}, plotted as a function of lookback time for a random subsample of 20 galaxies. The line colour corresponds to the stellar mass of each galaxy. A flaw in the fitting process can be observed at a lookback time of $10^{8.15} \: \rm yr$ -- namely that \texttt{STARLIGHT} is drawn to the lower bound of the stellar metallicity parameter space for this age bin. It is for this reason that we choose to exclude the SSP template with the lowest stellar metallicity value (i.e. $[\rm M / H]_{*} = -1.71$) at this specific age bin during the fitting process used to determine stellar metallicity histories in this work.}
    \label{fig:Metallicity_age_problem}
\end{figure}

The steps described above allow the stellar mass history of each galaxy to be ascertained at each of the 15 template ages. The SFH of the galaxies can also be determined at each age bin by finding the difference in stellar mass between two ages, and then dividing this mass difference by the time span between bins. Finally, using a similar procedure, we can alternatively collapse the two-dimensional parameter space in the other dimension to form a one-dimensional function of mass-weighted stellar metallicity, which we use to obtain the mean value for the mass-weighted stellar metallicity history for each galaxy at the 15 template ages.

\begin{figure}
	\includegraphics[width=0.47\textwidth]{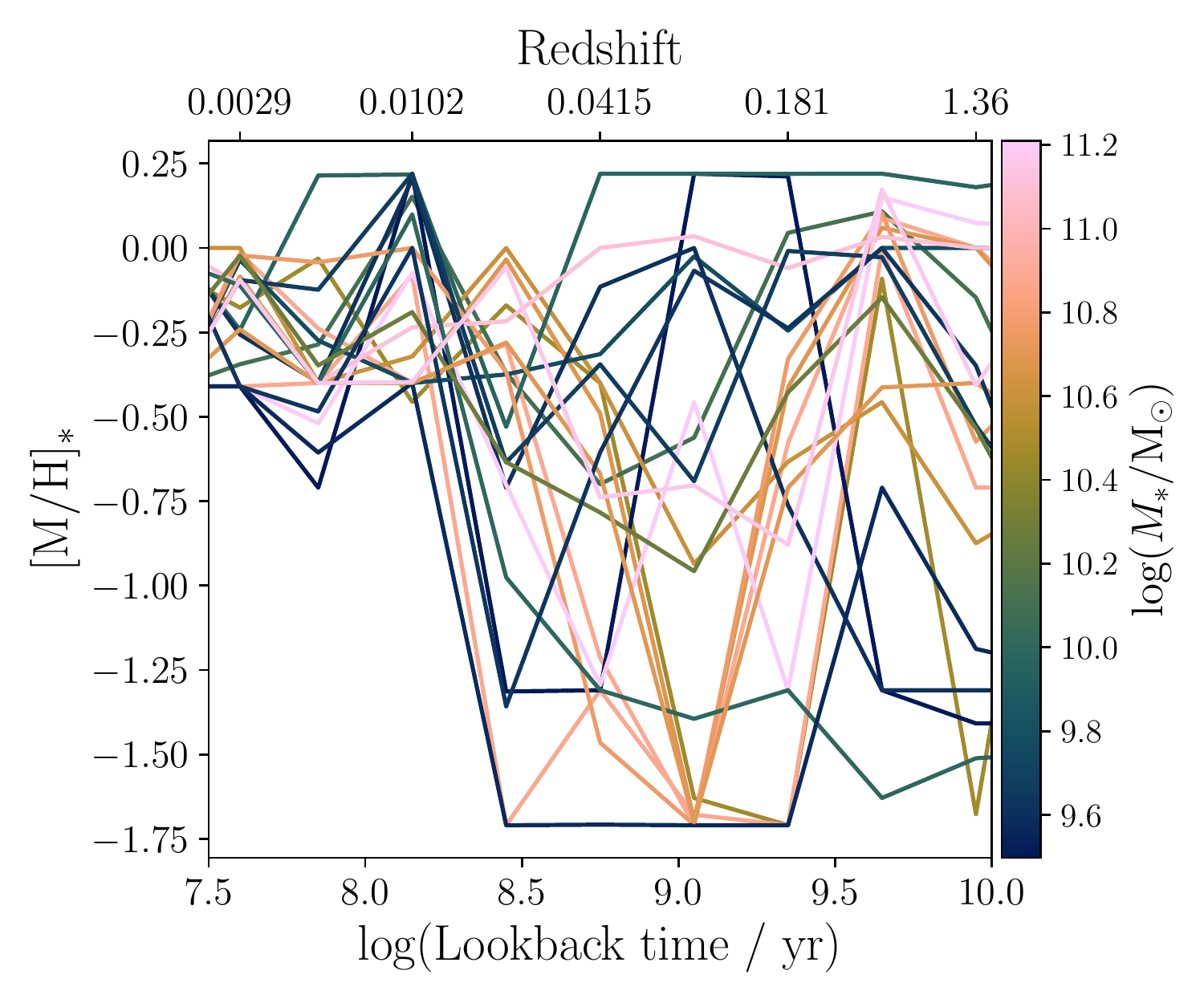}
    \caption{As Fig.~\ref{fig:Metallicity_age_problem} for the same random subsample of 20 galaxies, except now the problematic SSP template at $10^{8.15} \: \rm yr$ and $[\rm M / H]_{*} = -1.71$ is ignored during the \texttt{STARLIGHT} fitting process.}
    \label{fig:Metallicity_age_fixed}
\end{figure}

Since the galaxies which comprise the final sample used in this work are drawn from both the MaNGA Primary+ and Secondary samples (which obtain radial coverage out to $1.5 \: \rm R_{e}$ and $2.5 \: \rm R_{e}$ respectively; \citealp{Bundy2015OVERVIEWOBSERVATORY, Yan2016SDSS-IVQuality}), it is necessary to ensure that median stellar metallicities for galaxies in each of these samples can be compared with each other. Fortunately, the median value for the mass-weighted mean stellar metallicities for the Primary+ sample is $[\rm M / H]_{*} = -0.143$, while the corresponding value for the Secondary sample is $[\rm M / H]_{*} = -0.135$. Since there is not a substantial difference between these values, it is unlikely that the different radial coverage in the two samples has a significant influence on the results presented in this paper.

\medskip

It is well known that some of the SSP templates are ill-constrained by the fitting process \citep{Vazdekis2010EvolutionarySystem, Vazdekis2016UV-extendedGalaxies}. In particular, low stellar metallicity templates with ages around ${\sim} 10^{8} \: \rm yr$ are likely to fit systematic errors, such as residual flux calibration errors, rather than the data.

It is particularly difficult to produce reliable SSP templates for low metallicity populations. Creating these templates for populations that are dominated by intermediate age type B and A stars poses a challenge, since these stars are not only rare, but also contain very few metal lines in their spectra. Arguably, it is harder to create templates for these populations than those dominated by type O stars, since type B and A stars are vastly less luminous than their younger counterparts and hence their spectra are harder to observe. Like type O stars, the absence of metal lines makes the optical spectra of intermediate age populations poor stellar metallicity diagnostics. In fact, \citet{Vazdekis2010EvolutionarySystem, Vazdekis2016UV-extendedGalaxies} note that the templates of such stars flirt with being unreliable, as can be seen in Fig.~6 of \citet{Vazdekis2010EvolutionarySystem} and Figs.~3 and 5 of \citet{Vazdekis2016UV-extendedGalaxies}. Both Fig.~6 of \citet{Vazdekis2010EvolutionarySystem} and Fig.~3 of \citet{Vazdekis2016UV-extendedGalaxies} show that the threshold age at which the quality of the SSP templates at $[\rm M / H]_{*} = -1.71$ drops into the ``unsafe'' range is almost exactly $10^{8.15} \: \rm yr$. Indeed, if we perform a fit to the spectra with all of the templates at our disposal (Fig.~\ref{fig:Metallicity_age_problem}), we find that the lowest stellar metallicity template at a lookback time of $10^{8.15} \: \rm yr$ is hugely over-represented, leading to a systematic distortion in the stellar metallicity history of many galaxies. Such a drop is unphysical, and arises due to \texttt{STARLIGHT} latching on to the lowest stellar metallicity SSP template at this age, as this template is erroneously identified as the best fit for the galaxy. For an in-depth discussion about the reliability of the SSP templates used in this work, we refer the reader to Section~3.2 of \citet{Vazdekis2010EvolutionarySystem}, as well as Section~2 of \citet{Vazdekis2016UV-extendedGalaxies}.

The simplest course of action to try and address this issue is to exclude this lone problematic template at $10^{8.15} \: \rm yr$ prior to analysis. Throwing out this template does largely make the problem vanish, as can be seen in Fig.~\ref{fig:Metallicity_age_fixed}. While there may still be some residual issues at neighbouring ages, the stellar metallicities for both young ($\rm age < 10^{8.15} \: \rm yr$) and old ($\rm age > 10^{9.35} \: \rm yr$) populations do not appear to be at all affected by the removal of the SSP template at $10^{8.15} \: \rm yr$. Fortunately, since the young and old populations are not strongly coupled, they are robust against this potential systematic uncertainty -- particularly when compared to the templates of intermediate age stars. Finally, it is encouraging to see that the mass-weighted mean stellar metallicities in both Fig.~\ref{fig:Metallicity_age_problem} and Fig.~\ref{fig:Metallicity_age_fixed} converge to comparable, slightly sub-solar values at more recent lookback times. Despite our earlier concerns about whether the spectra of stars at such ages constitute reliable stellar metallicity diagnostics, this convergence means that we can now with confidence analyse mass-weighted mean stellar metallicities for populations dominated by stars in the range $10^{7.6} \: \rm yr \leq \rm age < 10^{8.15} \: \rm yr$.

In short, whether or not this subset of troublesome SSPs is included in the fit makes essentially no difference to the results obtained in the age intervals considered in this work. Although some systematic residual effects are still apparent in Fig.~\ref{fig:Metallicity_age_fixed}, the worst of them have been mitigated. Comparing Figs.~\ref{fig:Metallicity_age_problem}~and~\ref{fig:Metallicity_age_fixed}, we see that ages $< 10^{8.15} \: \rm yr$ and $> 10^{9.35} \: \rm yr$ are largely unaffected by these issues at intermediate ages. Accordingly, we can, with some confidence, compare the inferred mass-weighted mean stellar metallicities of galaxies at an age of $\rm 10^{9.95} \: \rm yr$ to those at the present day ($\rm age = 10^{7.6} \: \rm yr$). This allows us to robustly study chemical evolution since cosmic noon. For a detailed description of some of the tests we performed to check that the fitting procedure undertaken in this work is sound, we refer the interested reader to Appendix~\ref{sec:Appendix_A}.

\subsection{Gas Metallicities}
\label{subsec:Gas}

The gas metallicities of the galaxies are determined in two separate ways in this work. The first -- and most obvious -- method by which gas metallicities are calculated is by measuring the strengths of various emission lines in the galaxies' spectra. While this method is reasonably reliable\footnote{Calibration methods which rely on converting certain strong emission lines into gas metallicities often notoriously give systematically different values for the gas metallicity, depending on which calibration is used \citep[e.g.][]{Kewley2008MetallicityGalaxies, Maiolino2008AMAZE:3}. As \citet{Maiolino2019DeGalaxies} acknowledge, these so-called strong-line gas metallicity calibrations provide an ``easier, albeit less precise'' method by which gas metallicities may be determined.}, measuring the emission lines in these galaxies will naturally only allow the gas metallicities at the present day to be determined. In order to determine the gas metallicities of the galaxies at earlier times, we also employ the FMR of \citet{Mannucci2010AGalaxies}. These two separate methods are expanded upon below.

\subsubsection{Emission Lines}
\label{subsubsec:Emission_lines}

Although it is possible to directly measure the gas metallicity of a galaxy by measuring the strength of its ${\rm [\textsc{O~iii}]} \: \lambda 4363$ emission line, this line is very weak even in the best quality data and with the most appropriate chemical composition \citep[e.g.][]{Kewley2008MetallicityGalaxies, Kewley2019UnderstandingLines}. Furthermore, this so-called ``direct method'' may underestimate gas metallicities in both metal-rich galaxies \citep[e.g.][]{Stasinska2002TheNebulae, Stasinska2005BiasesNebulae, Bresolin2006TheAbundances} as well as metal-poor galaxies \citep[e.g.][]{Kobulnicky1999Chemical0.5}.

For these reasons, we instead choose to estimate gas metallicities in this work via the use of strong-line methods. Many different theoretical calibrations have been developed to convert certain emission-line ratios, known to be sensitive to gas metallicities, into gas metallicity estimates. However, in order to obtain gas metallicities from the multitude of strong line calibrations available, these various diagnostics have to first be converted to a common calibration scale. Fortunately, \citet{Kewley2008MetallicityGalaxies} find that it is possible to convert a given gas metallicity diagnostic into any other calibration scheme via:

\begin{equation}
    \log y = \sum_{n} c_{n} x^{n},
	\label{eq:calibration_conversion}
\end{equation} 
where $y$ is the original gas metallicity diagnostic to be converted in $12 + \log(\rm O / H)$ units, the $c_{n}$ are $n$th-order polynomial coefficients to be optimised, and $x$ is the gas metallicity (specifically, the oxygen abundance) relative to solar metallicity (which we assume to be $12 + \log(\rm O / H)_{\odot} = 8.69$; \citealp{Asplund2009TheSun}). For the purposes of subsequent comparison with average stellar metallicities, we convert gas metallicities derived from emission line measurements from their $12 + \log(\rm O / H)$ values to gas metallicity $\rm [M / H]_{g}$ via:

\begin{equation}
    \rm [M / H]_{g} = \log \left( Z_{g} / Z_{\odot} \right) = 12 + \log(\rm O / H) - 8.69,
	\label{eq:log_O_H_to_Z}
\end{equation}
\citep[e.g.][]{Sanders2021The3.3}. Here, $Z_{\odot}$ is the Solar metallicity value ($Z_{\odot} = 0.0142$; \citealp{Asplund2009TheSun}).

There are many available strong line calibrations: see, for instance, \citet{Kewley2008MetallicityGalaxies} for a comprehensive study comparing various different strong line calibrations, as well as to \citet{Scudder2021ConversionsMaNGA} for a similar, more recent study exclusive to MaNGA galaxies. However, according to both \citet{Kewley2008MetallicityGalaxies} and \citet{Kewley2019UnderstandingLines}, the $\rm N2O2$ ratio -- defined as $\rm [\textsc{N~ii}] \: \lambda 6584 / [\textsc{O~ii}] \: \lambda \lambda 3727, \: 3729$ -- is by far the most reliable optical gas metallicity diagnostic. Not only is it robust \citep{Paalvast2017MetallicityIMF}, but it also does not depend on the ionisation parameter \citep[e.g.][]{Kewley2002UsingGalaxies, Blanc2015IZI:Statistics}. The $\rm N2O2$ ratio is highly sensitive to gas metallicity for two reasons: firstly, because nitrogen may be formed by both primary and secondary nucleosynthesis processes \citep[e.g.][]{Considere2000StarburstsGradients, Kewley2019UnderstandingLines}; secondly, because the $[\textsc{O~ii}] \: \lambda \lambda 3727, \: 3729$ line is very sensitive to electron temperature \citep[e.g.][]{Hagele2008PrecisionGalaxies, Kewley2019UnderstandingLines}. The $\rm N2O2$ ratio depends on ISM pressure only at very high gas metallicities $\left[ 12 + \log(\rm O / H) > 9.23 \right]$ and at the highest ISM pressures \citep{Kewley2019UnderstandingLines}. Finally, the $\rm N2O2$ ratio is also the least sensitive optical diagnostic to the presence of an active galactic nucleus \citep{Kewley2006MetallicityFlows} or diffuse ionised gas \citep{Zhang2017SDSS-IVMeasurements}.

While gas metallicities are derived from oxygen abundances (cf. Equation~\ref{eq:log_O_H_to_Z}), stellar metallicities, by contrast, trace the iron abundance of stars \citep[e.g.][]{Tinsley1980a, Maiolino2019DeGalaxies}. Therefore, in order to treat the metallicities of the stellar populations and the gas from which these formed on equal footing, we must transform our elemental abundance base. Fortunately, \citet{Fraser-McKelvie2022TheGalaxies} have demonstrated that it is possible to convert gas metallicities into an iron base using the scaling relations of \citet{Nicholls2017AbundanceGalaxies} as follows:

\begin{equation}
\begin{aligned}
    {\rm [M / H]_{g, \: iron}} & = 0.9941 \times (12 + \log(\rm O / H)) - 8.9011, \\
                               & \ \ \ (\rm if \ 12 + \log(\rm O / H) - 8.69 > -0.5), \\
                               & = 0.6753 \times (12 + \log(\rm O / H)) - 8.6875, \\
                               & \ \ \ (\rm if \ 12 + \log(\rm O / H) - 8.69 < -0.5), \\
	\label{eq:gas_iron_base}
\end{aligned}
\end{equation} 
(A. Fraser-McKelvie, \emph{private communication}). Throughout the rest of this paper, any $\rm [M / H]_{g}$ values quoted will be understood to be scaled to an iron base, unless otherwise stated.

In this work, we use the best-fit coefficients calculated by \citet{Kewley2002UsingGalaxies}\footnote{$c_{0} = 1.54020$, $c_{1} = 1.26602$, and $c_{2} = 0.167977$.} in Equation~(\ref{eq:calibration_conversion}) to estimate the present-day gas metallicities of the MaNGA galaxies from measurements of the $\rm N2O2$ line ratio. Since -- as \citet{Kewley2008MetallicityGalaxies} caution -- the absolute gas metallicity determined from any particular strong-line method should not be trusted, the final step is to calibrate the gas metallicities calculated in this work to the stellar mass--gas metallicity relation ($\rm MZ_{g}R$) of \citet{Tremonti2004TheSDSS}. This is done by forcing the best fit line to our data to have the same value for the gas metallicity predicted by the $\rm MZ_{g}R$ of \citet{Tremonti2004TheSDSS} at a stellar mass of $10^{10} \ \rm M_{\odot}$. Doing so decreases the calculated gas metallicities by $0.3 \: \rm dex$. Note that the $\rm MZ_{g}R$ is distinct from the stellar mass--stellar metallicity relation, which we denote $\rm MZ_{*}R$ throughout this paper for the sake of clarity.

\subsubsection{The Fundamental Metallicity Relation}
\label{subsubsec:FMR}

Since the aim of this paper is to compare the chemical evolution of the stars and the gas within spiral galaxies, we require an alternative method for determining the gas metallicities at epochs earlier than the present day. To this end, we turn to the FMR obtained by \citet{Mannucci2010AGalaxies}, who demonstrate that the gas metallicity of a galaxy may be estimated if both its stellar mass, $M_{*}$, and its SFR are known:

\begin{figure*}
	\includegraphics[width=0.95\textwidth]{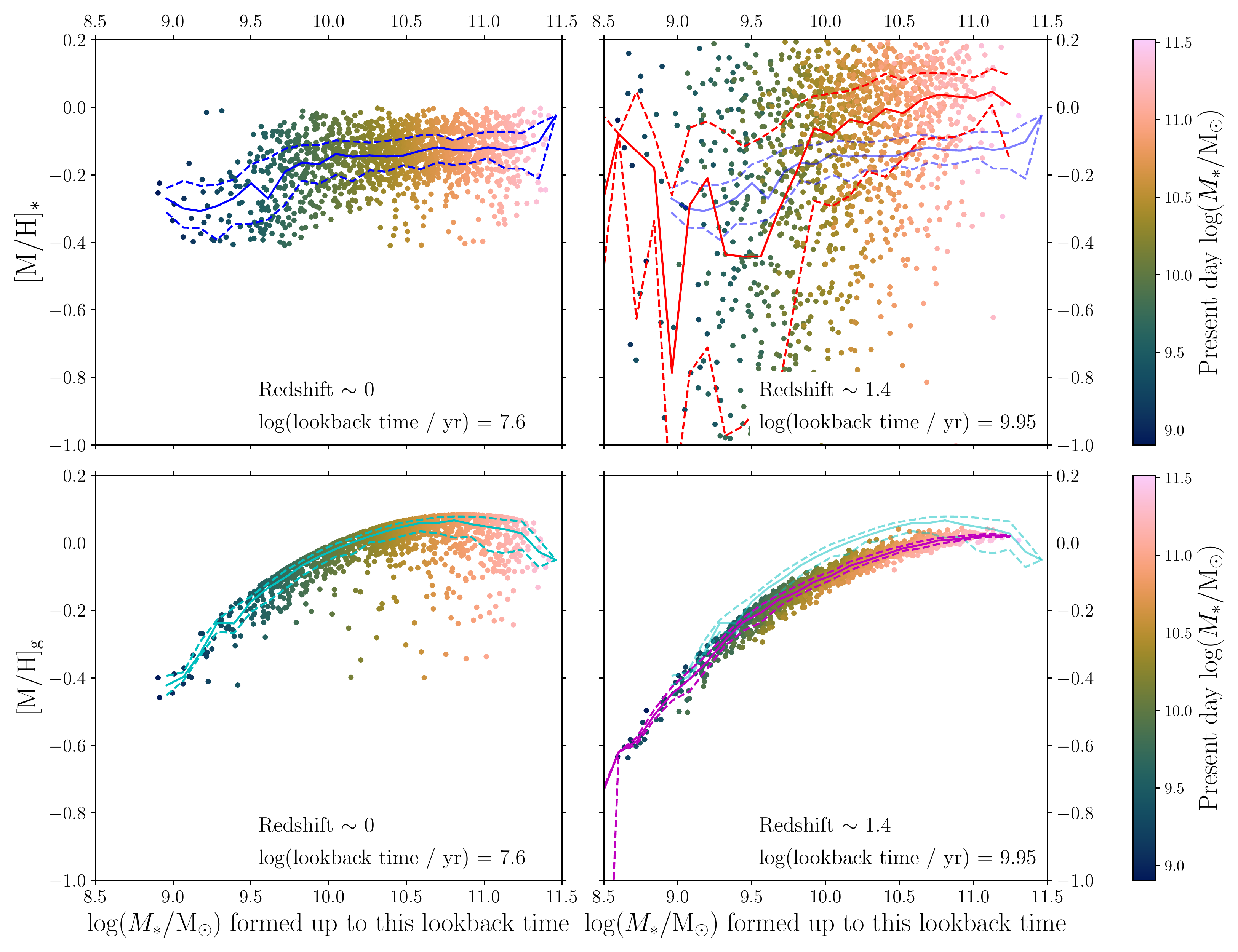}
    \caption{Mass--metallicity relations for the sample of MaNGA galaxies. Individual galaxies are coloured by their present-day stellar masses. The horizontal axis shows the mass in stars that has formed up to a certain lookback time. The plots on the top row show mass-weighted mean stellar metallicity as a function of stellar mass, while the plots on the bottom row show gas metallicity as a function of stellar mass. The data in the left-hand column are from the youngest age bin at a lookback time of $10^{7.6} \: \rm yr$, while the right-hand column shows data from the oldest age bin at a lookback time of $10^{9.95} \: \rm yr$. The running medians (solid lines) and running inter-quartile ranges (dashed lines) to the mass-weighted mean stellar metallicity data at high and low redshift are overplotted in red and blue, respectively. Similarly, the equivalent lines for the gas metallicities at high and low redshift are overplotted in magenta and cyan, respectively.}
    \label{fig:MZR_plots}
\end{figure*}

\begin{equation}
\begin{aligned}
    12 + \log(\rm O / H) = & \ 8.90 + 0.37 m - 0.14 s - 0.19 m^{2} \\
                         & + 0.12 ms - 0.054 s^{2}.
	\label{eq:FMR}
\end{aligned}
\end{equation} 
Here, $m = \log \left( M_{*} \right) - 10$, with $M_{*}$ measured in units of $\rm M_\odot$; and $s = \log \left( \rm SFR \right)$, with SFR measured in units of $\rm M_\odot \ yr^{-1}$. \citet{Mannucci2010AGalaxies} find that the FMR does not evolve until a redshift of $z \sim 2.5$, while \citet{Sanders2021The3.3} conclude that there is no evolution in the FMR until $z \sim 3.3$. Therefore, we can use Equation~(\ref{eq:FMR}) to comfortably estimate the gas metallicities of galaxies as far back as the largest lookback time we consider in this work, $10^{9.95} \: \rm yr$ ($z \sim 1.4$). Again, we convert all gas metallicities determined by using the FMR into $\rm [M / H]_{g}$ values, before calibrating the present-day values against the $\rm MZ_{g}R$ calculated by \citet{Tremonti2004TheSDSS}, as described in Section~\ref{subsubsec:Emission_lines}. This calibration results in a decrease in $\rm [M / H]_{g}$ of $0.31 \: \rm dex$ at the present day.

Reassuringly, as we shall see from the analysis in Section~\ref{subsec:mass_dependency_metallicity}, gas metallicities inferred from the FMR at recent epochs agree very well with those obtained directly from the gas itself via measurements of the N2O2 line ratio.

\section{Results and Discussion}
\label{sec:Results}

We first plot the mass--metallicity relations for both the stellar populations and the gaseous ISM at low and high redshifts. This also allows us to determine how both the $\rm MZ_{*}R$ and the $\rm MZ_{g}R$ of the sample galaxies evolve over cosmic time. We then explore the ramifications of these results further by investigating how stellar mass influences the evolution of the average stellar and gas metallicities of these galaxies.

\subsection{Evolution of the Mass--Metallicity Relations}
\label{subsec:MZR_comparison}

Figure~\ref{fig:MZR_plots} shows mass--metallicity relations for the sample of 1619 MaNGA galaxies. Plots on the top row show the mass-weighted mean stellar metallicities for each of the sample galaxies, and plots on the bottom row show gas metallicity data. The plots in the left-hand column show the $\rm MZ_{*}R$ and the $\rm MZ_{g}R$ at the closest lookback time to the present day, $10^{7.6} \: \rm yr$ ($z \sim 0$), while those in the right-hand column show how both the mass--metallicity relations were at a lookback time of $10^{9.95} \: \rm yr$ ($z \sim 1.4$).

There are two kinds of uncertainties that we have to consider in Fig.~\ref{fig:MZR_plots}. The first kind are the random errors from the scatter in the data used to construct the plots of Fig.~\ref{fig:MZR_plots}. The second kind of uncertainties are systematic errors in the $\rm MZ_{*}R$. Although these errors are fundamentally unquantifiable, fortunately they will at least all be internally consistent at all redshifts, since we have determined the $\rm MZ_{*}R$ in exactly the same way both at low and high redshift.

\medskip

The gas metallicities exhibit a much tighter relationship with stellar mass than do the mass-weighted mean stellar metallicities. This is because they are estimated (via the FMR -- see Equation~\ref{eq:FMR}) using just two input parameters: the stellar mass and the SFR at a particular age. This means that the tight relationship in the $\rm MZ_{g}R$ is actually a consequence of the tight relationship between stellar mass and SFR. In order to test how reliably the FMR reproduces the $\rm MZ_{g}R$, we investigated at both low and high redshift the relationship between stellar mass and SFR. Reassuringly, the evolution in this parameter space from high to low redshift was found to be in good agreement with that reported by \citet{Speagle2014A0-6}. Additionally, to further verify that the FMR robustly reproduces the $\rm MZ_{g}R$ at low redshift, we also independently derived the $\rm MZ_{g}R$ at $z \sim 0$ using the N2O2 calibrator of \citet{Kewley2002UsingGalaxies} -- see Section~\ref{subsubsec:Emission_lines} -- and found that these two different methods yielded consistent relationships between stellar mass and gas metallicity

Figure~\ref{fig:MZR_plots} shows that the evolution of the $\rm MZ_{g}R$ between $z \sim 1.4$ and $z \sim 0$ is marginal; nevertheless, the $\rm MZ_{g}R$ is seen to increase to higher average gas metallicity values between a redshift of $z \sim 1.4$ and the present day. This result is in line with many authors who have previously studied the redshift evolution of the $\rm MZ_{g}R$ -- for instance \citet{Maiolino2008AMAZE:3, Moustakas2011EvolutionZ=0.75, Zahid2013TheYears, Zahid2014TheRelation, Lian2018TheIMF, Lian2018Modelling0}; and \citet{Yates2021L-GALAXIESGalaxies}. In particular, the $\rm MZ_{g}R$ at $z \sim 0$ shown in Fig.~\ref{fig:MZR_plots} is in very good agreement with the results reported by both \citet{Fontanot2021TheModel} and \citet{Curti2020TheGalaxies}, the latter of which find a peak O/H value of ${\sim} 0.1$ dex above Solar, just as we report here. While our $\rm MZ_{g}R$ at $z \sim 1.4$ is systematically lower than that reported by \citet{Fontanot2021TheModel}, it is still within their range of uncertainty (see also the large spread of observational results from different authors such as \citealp{Zahid2014TheRelation}, \citealp{Yabe2015The1.4}, and \citealp{Curti2020TheGalaxies}).

A similar evolution in the $\rm MZ_{*}R$ has been found from semi-analytic modelling performed by \citet{Yates2021L-GALAXIESGalaxies}, as well as at a high redshift range of $1.6 \leq z \leq 3.0$ by \citet{Kashino2022Thealpha-enhancement}. Observational results at low redshifts from authors such as \citet{Camps-Farina2021EvolutionGalaxies, Camps-Farina2022ChemicalGalaxies} and \citet{Fontanot2021TheModel} also show that the $\rm MZ_{*}R$ does indeed evolve from high to low redshift, with high-mass galaxies found to evolve faster than low-mass ones. It is important to note, however, that \citet{Camps-Farina2021EvolutionGalaxies, Camps-Farina2022ChemicalGalaxies} also find that the chemical evolution history of a galaxy may be strongly influenced by other fundamental properties, such as their stellar masses and morphologies. In addition, \citet{Beverage2021ElementalQuenching} find that the $\rm MZ_{*}R$ may not evolve with redshift at all, depending on the elemental abundance concerned.

It can be seen from Fig.~\ref{fig:MZR_plots} that there has been little evolution in the median $\rm MZ_{*}R$ with redshift since a lookback time of ${\sim} 10^{9.95} \: \rm yr$. While a similar lack of evolution in the $\rm MZ_{*}R$ over this time period is reported by \citet{Panter2008TheRecord} and \citet{ValeAsari2009TheAstropaleontology}, other authors, such as \citet{Camps-Farina2021EvolutionGalaxies, Camps-Farina2022ChemicalGalaxies} show that $[\rm M / H]_{*}$ has decreased since this lookback time. However, the overall spread of the $\rm MZ_{*}R$ clearly does evolve with redshift. Fundamentally, there is a very large spread in the observed mass-weighted mean stellar metallicities of galaxies, particularly at higher redshifts. This can be seen in this work in the top-right plot of Fig.~\ref{fig:MZR_plots}, as well as in work by previous authors such as \citet{Gallazzi2005TheUniverse, Panter2008TheRecord}; and \citet{Beverage2021ElementalQuenching}. The spread of the data in this top-right plot is very similar to the $\rm MZ_{*}R$ found by \citet{Panter2008TheRecord}. By contrast, the spread of the data in the low-redshift $\rm MZ_{*}R$ is much tighter, a result also reported at low redshift by \citet{Fontanot2021TheModel} -- though note the spread of the data in our high-redshift $\rm MZ_{*}R$ is larger than that reported at high redshift by these same authors. The range of our average stellar metallicities in the top-left plot of Fig.~\ref{fig:MZR_plots} is between $-0.4$ and $0.0$ dex, and the median $\rm MZ_{*}R$ reaches roughly solar abundance at the very high-mass end, in agreement with previous authors including \citet{Zahid2017StellarRate} and \citet{Fontanot2021TheModel}.

In essence, as we move closer to the present day, the mass-weighted mean stellar metallicities exhibit a much tighter relation with stellar mass, and are comparable to the observed gas metallicities; the data here are actually in much better agreement with the $\rm MZ_{g}R$ of \citet{Tremonti2004TheSDSS}. Such a finding is in accord with previous work by authors such as \citet{GonzalezDelgado2014InsightsSurvey}, \citet{Sanchez2018SdssGalaxies}, and \citet{Lacerda2020GalaxiesSurvey}, each of whom find good agreement between the $\rm MZ_{g}R$ and the $\rm MZ_{*}R$ at young stellar ages. This makes sense when we recall that this is the gas from which these same stars are forming. That the mass-weighted mean stellar metallicities converge in this manner over cosmic time is not a consequence of our sample selection; in reality it is a consequence of the rich and varied chemical evolution histories of the individual galaxies over several generations of star formation. As \citet{Fraser-McKelvie2022TheGalaxies} explain, the difference between stellar and gas metallicities at early cosmic times is dependent on the SFH of the galaxy. Galaxies that reach their peak SFR at later times take longer to accumulate metals, and so $[\rm M / H]_{*} \ll [\rm M / H]_{g}$ \citep{Fraser-McKelvie2022TheGalaxies}.

\medskip

The analysis presented in this work ultimately represents a step forward over complementary methodology by which gas metallicities are also observationally determined -- i.e. obtaining snapshots of the metallicity of the gas in various galaxies at increasing redshifts. While the quality of data obtained via these snapshots decreases with increasing redshift, this alternative method of galactic archaeology, by contrast, acquires high-quality measurements of the gas metallicity back to distant lookback times, as seen in Fig.~\ref{fig:MZR_plots}. Such analysis is similar to that which has already been carried out for the stellar metallicity histories of galaxies by authors such as \citet{Peterken2019Time-slicingMaNGA, Peterken2020SDSS-IVGalaxies}, \citet{Camps-Farina2021EvolutionGalaxies, Camps-Farina2022ChemicalGalaxies}, and \citet{Fraser-McKelvie2022TheGalaxies}. Furthermore, since we are tracking the evolution of the gas (as well as the stellar populations) from observations made in very low-redshift galaxies, we are also able to check the validity of our archaeological approach against present-day emission-line data in these same galaxies. This concept is explored further in Section~\ref{subsec:mass_dependency_metallicity}.

\subsection{How does Stellar Mass Affect Metallicity Evolution?}
\label{subsec:mass_dependency_metallicity}

\begin{figure}
	\includegraphics[width=0.47\textwidth]{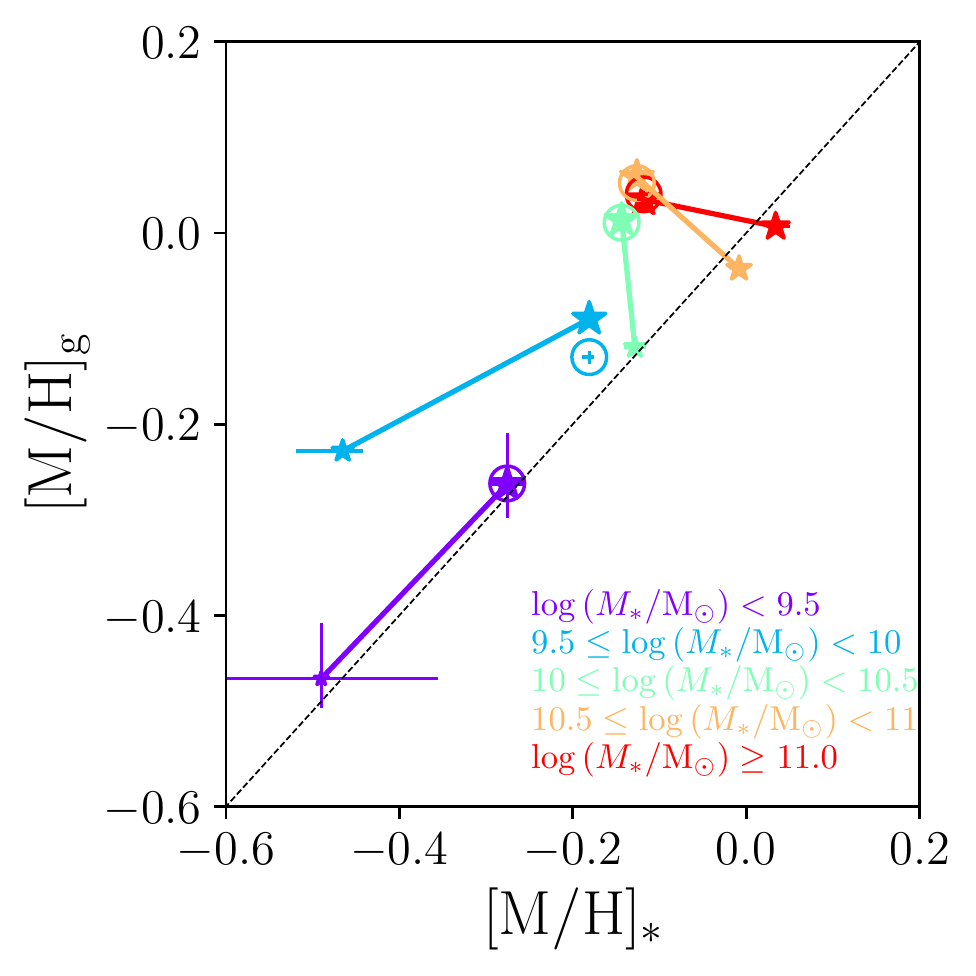}
    \caption{Evolution of stellar and gas metallicities with cosmic time. The median metallicities for the sample galaxies are determined for five different stellar mass bins. Three datapoints are plotted for each mass bin. Star-shaped datapoints denote mass-weighted mean stellar metallicity data derived from \texttt{STARLIGHT} and gas metallicity data from the FMR of \citet{Mannucci2010AGalaxies}. The size of the star is proportional to the fraction of mass that had formed at that redshift. The larger stars therefore correspond to a lookback time of $10^{7.6} \: \rm yr$ ($z \sim 0$), and the smaller stars to a lookback time of $10^{9.95} \: \rm yr$ ($z \sim 1.4$). The open circles plot the stellar metallicities of the mass bins at a lookback time of $10^{7.6} \: \rm yr$ against the present-day gas metallicities derived from measurements of the N2O2 ratio \citep{Kewley2002UsingGalaxies} to ensure the gas metallicities predicted by the FMR are plausible. The errorbars show the $1 \sigma$ uncertainty on the median of each datapoint. The black diagonal line is a 1-to-1 relation, and does not represent a fit to the data.}
    \label{fig:Metallicity_histories}
\end{figure}

In order to investigate how stellar mass influences metallicity evolution, we now turn to Fig.~\ref{fig:Metallicity_histories}, which shows how both mass-weighted mean stellar metallicity, $[\rm M / H]_{*}$, and gas metallicity, $[\rm M / H]_{g}$, have evolved with cosmic time since cosmic noon. We have divided the galaxies into five logarithmically-spaced stellar mass bins, and calculated the median of the gas and mass-weighted mean stellar metallicity for each mass bin. The star-shaped datapoints are generated using mass-weighted mean stellar metallicity data from \texttt{STARLIGHT}, and gas metallicity data from the FMR of \citet{Mannucci2010AGalaxies}; these evolve from a lookback time of $10^{9.95} \: \rm yr$ through to $10^{7.6} \: \rm yr$. The open circles show -- as a comparison -- the same stellar metallicities at a lookback time of $10^{7.6} \: \rm yr$, but plotted instead against the present-day gas metallicities that have been directly determined from measurements of the N2O2 ratio. These emission line data are included only as a consistency check to ensure that the data derived from the FMR (at the present day) are sensible. The two complementary methods for determining gas metallicities at the present epoch yield consistent values for galaxies in all mass bins. To summarise, Fig.~\ref{fig:Metallicity_histories} represents the ultimate result of the co-evolution of gas and stellar metallicity that we originally set out to measure. As mentioned in Section~\ref{subsubsec:FMR}, the good agreement between the present-day gas metallicity derived from the FMR and that measured directly from emission lines gives some confidence in the validity of this approach.

\medskip

In interpreting these results, let us first consider those spiral galaxies with stellar masses less than $10^{10} \ \rm M_{\odot}$. Generically, these galaxies are seen to increase both their stellar and gas metallicities since cosmic noon. The increases in the median gas metallicities are consistent with those measured by \citet{Panter2008TheRecord} and \citet{Maiolino2008AMAZE:3}, and similar increases in stellar metallicities over this time period are also seen by \citet{Camps-Farina2021EvolutionGalaxies, Camps-Farina2022ChemicalGalaxies}. While certain authors, such as \citet{SanchezAlmeida2017GasRates}, \citet{Sanchez-Menguiano2019CharacterizingGalaxies}, and \citet{Belfiore2019FromGradients}, have reported that gas accretion may in fact play an important role in galactic evolution, even in low-mass systems, the findings presented in this paper fit in well with the results from recent work (\citealp{Greener2021SDSS-IVRevisited}; see also \citealp{Mejia-Narvaez2020TheSurvey} and the semi-analytic spectral fitting performed by \citealp{Zhou2022Semi-analyticGalaxies}), in which we conclude that such relatively low-mass spiral galaxies tend to evolve as closed boxes (i.e. very little or no gas flows into or out of the galaxy over the course of its life -- see \citealp{Talbot1971TheModel} and \citealp{Tinsley1974ConstraintsNeighborhood}).

Within the context of these results, the behaviour of the low-mass spiral galaxies in Fig.~\ref{fig:Metallicity_histories} is to be anticipated. Since these low-mass galaxies evolve as closed boxes but are still forming stars at the present epoch, they must be forming stars over much longer timescales than their high-mass counterparts. Previous authors, such as \citet{ValeAsari2007TheSurvey}, have attributed the languid star formation within such galaxies to their proportionally lower gas metallicities (see Fig.~\ref{fig:Metallicity_histories}). However, it is more likely that the low stellar masses of such galaxies are actually the driving force behind their lengthy star formation timescales (\citealp{Greener2021SDSS-IVRevisited, Peterken2021SizeMaNGA}; see also \citealp{Camps-Farina2021EvolutionGalaxies, Camps-Farina2022ChemicalGalaxies} and \citealp{Fraser-McKelvie2022TheGalaxies}), since stellar mass is the dominant factor influencing the gas metal content of galaxies \citep[e.g.][]{Tremonti2004TheSDSS}. This leisurely pace allows these low-mass galaxies to thoroughly mix their gas between stellar generations, and thus they steadily increase both the metallicity of their stars -- and the ISM which these stars pollute -- over cosmic time, exactly as we see in Fig.~\ref{fig:Metallicity_histories}.

\medskip

Figure~\ref{fig:Metallicity_histories} shows that high-mass galaxies have also increased their median gas metallicities since $z \sim 1.4$ (albeit less dramatically than their low-mass counterparts; again, this is consistent with the direct measurements made by \citealp{Panter2008TheRecord}, \citealp{Maiolino2008AMAZE:3}, and \citealp{Camps-Farina2021EvolutionGalaxies, Camps-Farina2022ChemicalGalaxies}). In \citet{Greener2021SDSS-IVRevisited}, we showed that a spiral galaxy with a stellar mass greater than $10^{10} \ \rm M_{\odot}$ is far more likely to behave instead as an accreting box rather than a closed box. The accreting box model is similar to that of a closed box, but additionally allows for a steady stream of pristine gas to flow into the galaxy over time \citep{Larson1972b, Larson1976a, Tinsley1974ConstraintsNeighborhood, Tinsley1980a}. Moreover, if galaxies do not accrete pristine gas, their star formation quenches rapidly; this effect is also far more pronounced for higher mass galaxies in the epoch since cosmic noon (E. Taylor et al. 2022, \emph{submitted}).

This infall of metal-poor gas allows these galaxies to produce a significant quantity of high-mass stars, which, upon their deaths, pollute the ISM with newly-forged metals. Subsequent generations of stars will be formed from this enriched material, which ultimately means that over cosmic time, the gas metallicity of high-mass galaxies gradually increases.

\medskip

We might expect that the stellar populations in high-mass galaxies should be similarly enriched over cosmic time; however, Fig.~\ref{fig:Metallicity_histories} demonstrates that the median stellar metallicity of such galaxies decreases at later times. This is a surprising result, and is in disagreement with other authors such as \citet{Panter2008TheRecord} -- who find essentially no change in the average stellar metallicity over this redshift interval -- and \citet{Camps-Farina2021EvolutionGalaxies, Camps-Farina2022ChemicalGalaxies} -- while these authors do report a decrease in the average stellar metallicity since ${\sim} 10^{9.75} \: \rm yr$, this decrease is less pronounced than that which we report here. Why should this be? High-mass galaxies must accrete largely unmixed pristine gas over their lifetimes in order to satiate the demand required to maintain star formation at later times. Furthermore, these galaxies have been found to form the majority of their stars at earlier times \citep[e.g.][]{Beverage2021ElementalQuenching, Peterken2021SizeMaNGA, Zhou2021StarUniverse, Fraser-McKelvie2022TheGalaxies}.

The decrease in the median stellar metallicities of these galaxies over cosmic time could, therefore, be a direct consequence of the pristine gas that they accrete. If this gas is not well mixed into the galaxy, it is probable that metal-poor stars will continuously form, even at very late times. These stars will still pollute the ISM with metals upon their deaths, raising the average metallicity of the gas in these galaxies over time; however, even at late times the average stellar metallicity may be suppressed if such metal-poor stars always form from the supply of pristine gas. Further work will be required to investigate whether the average stellar metallicities observed in high-mass galaxies in this work at the redshift of $z \sim 1.4$ around cosmic noon represents a local maximum, or whether the median stellar metallicities of such galaxies increases yet further at even higher redshifts.

\section{Conclusions}
\label{sec:Conclusions}

We investigate the chemical evolution histories of both the stellar populations and the gas for a well-defined sample of 1619 spiral galaxies observed by MaNGA. Mass-weighted mean stellar metallicities are determined at low ($z \sim 0$) and high ($z \sim 1.4$) redshifts via the use of the full-spectrum stellar population synthesis code \texttt{STARLIGHT} \citep{CidFernandes2005Semi-empiricalMethod}; gas metallicities are estimated at the same epochs using the FMR of \citet{Mannucci2010AGalaxies}. We also calculate the present-day gas metallicities of the same galaxies by using the strong-line N2O2 calibration proposed by \citet{Kewley2002UsingGalaxies}.

The evolution of both the $\rm MZ_{*}R$ and the $\rm MZ_{g}R$ are examined and discussed, in addition to the mass-dependence of the average stellar and gas metallicities of the galaxies. The results of these investigations are summarised below.

\begin{enumerate}

  \item The $\rm MZ_{g}R$ is generically found to evolve in such a way that at a given mass, the gas within the galaxies becomes increasingly metal-rich with cosmic time. This result is in agreement with the findings of \citet{Maiolino2008AMAZE:3}, \citet{Moustakas2011EvolutionZ=0.75}, \citet{Zahid2013TheYears, Zahid2014TheRelation}, \citet{Lian2018TheIMF, Lian2018Modelling0}, and \citet{Yates2021L-GALAXIESGalaxies}. Unlike the methods by which these authors determine the evolution of the $\rm MZ_{g}R$, however, the method used in this work of tracking the evolution of the gas within low-redshift galaxies back to more distant lookback times via galactic archaeology allows for high-quality measurements of the gas metallicity even at high redshifts.
    
  \vspace{1mm}
  
  \item The median $\rm MZ_{*}R$ exhibits a general decline in stellar metallicity from $z \sim 1.4$ to $z \sim 0$, for the bulk of the explored sample, with the possible exception for the mass range $M_{*} < 10^{10} \: \rm M_{\odot}$, in which little to no evolution is seen. The shape of the $\rm MZ_{*}R$ also changes: it is steeper at $z \sim 1.4$ and shallower at $z \sim 0$. The spread of the mass-weighted mean stellar metallicity data -- comparable to that found by \citet{Gallazzi2005TheUniverse}, \citet{Panter2008TheRecord}, and \citet{Beverage2021ElementalQuenching} -- also evolves with cosmic time: the $\rm MZ_{*}R$ exhibits a much tighter relation at later times. In fact, at $z \sim 0$, the observed $\rm MZ_{*}R$ is more comparable to the $\rm MZ_{g}R$ of \citet{Tremonti2004TheSDSS} than, for instance, the $\rm MZ_{*}R$ of \citet{Panter2008TheRecord}, since these stars are forming from the same gas whose emission lines authors such as \citet{Tremonti2004TheSDSS} observe.
  
  \vspace{1mm}
  
  \item The results in this paper align well with those found in recent work \citep{Greener2021SDSS-IVRevisited}, in which we conclude that low-mass spirals evolve as closed boxes, whereas high-mass spirals accrete a stream of relatively pristine gas over the course of their lives. We find in this work that low-mass galaxies steadily increase their stellar and gas metallicities over cosmic time since they form stars relatively slowly and are parsimonious with their gas reservoirs. High-mass galaxies, by contrast, have lower average stellar metallicities at later times. If the pristine gas is not particularly well-mixed into the galaxy upon accretion, large numbers of metal-poor stars will readily form in that galaxy even at very late times. This could result in the average stellar metallicities in high-mass spiral galaxies being suppressed at the present epoch.

\end{enumerate}

\section*{Data Availability}

This publication uses MPL-11 MaNGA science data products. The full sample of data used in this paper has recently been made completely accessible as part of the public release of SDSS DR17 \citep{Abdurrouf2022TheData}.

\section*{Acknowledgements}

We would like to thank our friend and colleague, Amelia Fraser-McKelvie, for many useful conversations and for very kindly helping to convert the gas metallicities in this work from an oxygen base into an iron base.

We are grateful to both Nicholas Fraser Boardman and Rog{\'e}rio Riffel for their myriad comments and suggestions which improved this manuscript.

Funding for the Sloan Digital Sky Survey IV has been provided by the Alfred P. Sloan Foundation, the U.S. Department of Energy Office of Science, and the Participating Institutions. SDSS acknowledges support and resources from the Center for High-Performance Computing at the University of Utah. The SDSS website is \url{www.sdss.org}.

SDSS is managed by the Astrophysical Research Consortium for the Participating Institutions of the SDSS Collaboration including the Brazilian Participation Group, the Carnegie Institution for Science, Carnegie Mellon University, the Chilean Participation Group, the French Participation Group, Harvard-Smithsonian Center for Astrophysics, Instituto de Astrof{\'i}sica de Canarias, The Johns Hopkins University, Kavli Institute for the Physics and Mathematics of the Universe (IPMU) / University of Tokyo, the Korean Participation Group, Lawrence Berkeley National Laboratory, Leibniz Institut f{\"u}r Astrophysik Potsdam (AIP), Max-Planck-Institut f{\"u}r Astronomie (MPIA Heidelberg), Max-Planck-Institut f{\"u}r Astrophysik (MPA Garching), Max-Planck-Institut f{\"u}r Extraterrestrische Physik (MPE), National Astronomical Observatories of China, New Mexico State University, New York University, University of Notre Dame, Observat{\'o}rio Nacional / MCTI, The Ohio State University, Pennsylvania State University, Shanghai Astronomical Observatory, United Kingdom Participation Group, Universidad Nacional Aut{\'o}noma de M{\'e}xico, University of Arizona, University of Colorado Boulder, University of Oxford, University of Portsmouth, University of Utah, University of Virginia, University of Washington, University of Wisconsin, Vanderbilt University, and Yale University.

This research made use of \texttt{Astropy} \citep{Robitaille2013Astropy:Astronomy}; \texttt{Marvin} \citep{Cherinka2018Marvin:Set}; \texttt{Matplotlib} \citep{Hunter2007b}; \texttt{NumPy} \citep{VanderWalt2011b}; \texttt{SciPy} \citep{Virtanen2020SciPyPython}; and \texttt{TOPCAT} \citep{Taylor2005b}.




\bibliographystyle{mnras}
\bibliography{main.bib} 



\appendix
\setcounter{section}{0}
\renewcommand{\thesection}{\Alph{section}}

\section{Additional Tests of the Spectral Fitting Outputs}
\label{sec:Appendix_A}
\setcounter{figure}{0}
\renewcommand{\thesubsection}{\Alph{section}\arabic{subsection}}
\renewcommand\thefigure{\Alph{section}\arabic{figure}}

In order to verify that the mass-weighted mean stellar metallicities and SFRs obtained from the \texttt{STARLIGHT} fitting procedure are reliable -- even after removing the problematic SSP template at $\log(\rm age / yr) = 8.15$ and $[\rm M / H]_{*} = -1.71$ prior to analysis (as described in detail in Section~\ref{subsec:Stellar}) -- we have undertaken several tests which we describe in detail below.

\subsection{Comparing Best-Fit Spectra}
\label{subsec:testing_SSPs}

\texttt{STARLIGHT} uses Monte Carlo techniques to derive best-fit spectra for the input spectra of each of the spaxels in the sample galaxies, with no assumptions made on the shape of the derived star formation histories \citep{CidFernandes2005Semi-empiricalMethod}. Each best-fit spectrum is created from a linear combination of a set of input SSP template spectra and an applied dust attenuation. We wish to test whether the resulting best-fit spectrum produced for a given galaxy is affected by the removal of a single SSP template at $\log(\rm age / yr) = 8.15$ and $[\rm M / H]_{*} = -1.71$.

\begin{figure*}
	\includegraphics[width=0.75\textwidth]{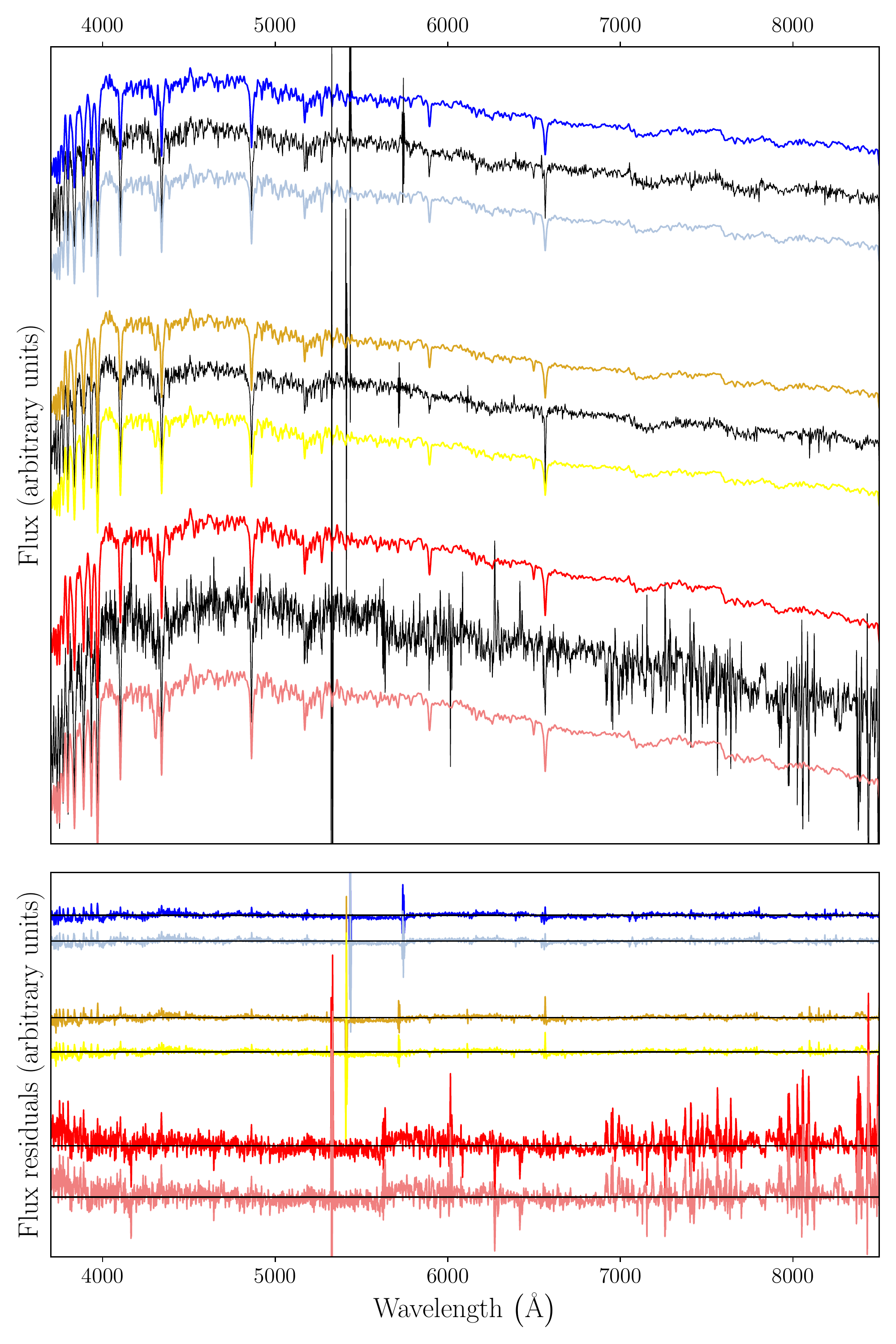}
    \caption{Example spectra (upper plot) and residuals (lower plot) for three randomly selected sample galaxies with varying S/N levels. Blue spectra are those of galaxy 9499-6102 ($\rm S / N = 93.4$); yellow spectra are those of galaxy 9024-6102 ($\rm S / N = 46.2$); red spectra are those of galaxy 8619-12703 ($\rm S / N = 26.2$). Black lines show the raw MaNGA spectra for each of these three galaxies. Light coloured lines (plotted below the corresponding black spectra) show the best-fit spectra (upper plot), and residuals between these spectra and the raw MaNGA spectra (lower plot), produced by \texttt{STARLIGHT} if all the SSP templates are included in the fitting procedure. Dark coloured lines (plotted above the corresponding black spectra) show the best-fit spectra and residuals if the problematic SSP template at $\log(\rm age / yr) = 8.15$ and $[\rm M / H]_{*} = -1.71$ is excluded from the fitting procedure (cf. Section~\ref{subsec:Stellar}). Note that the vertical axis, showing the flux of the stellar spectra, is arbitrary, since spectra are offset from each other by arbitrary amounts for the sake of clarity. Almost no differences can be seen between either any of the three pairs of best-fit spectra or any of their residuals.}
    \label{fig:SSP_spectra}
\end{figure*}

The upper plot of Fig.~\ref{fig:SSP_spectra} compares -- for each of three randomly selected sample galaxies of varying signal-to-noise ratio (S/N) levels -- the raw MaNGA spectrum (black line), the best-fit spectrum produced by \texttt{STARLIGHT} if all the SSP templates (described in Section~\ref{subsec:Stellar}) are included in the fitting procedure (light coloured line), and the equivalent best-fit spectrum if the problematic SSP template at $\log(\rm age / yr) = 8.15$ and $[\rm M / H]_{*} = -1.71$ is excluded prior to the fitting procedure (dark coloured line). For each of the three galaxies, the fluxes of the best-fit spectra have been arbitrarily offset to either side of the raw MaNGA spectrum so that each of the spectra are clearly visible; otherwise, both spectra would be plotted almost exactly on top of each other for a given galaxy.

The lower plot of Fig.~\ref{fig:SSP_spectra} shows -- for each of the three galaxies -- the flux residuals between the raw MaNGA spectrum and the two best-fit spectra described above. Again, each pair of flux residuals -- and their respective black baselines -- have been arbitrarily vertically offset from each other, so that both can be clearly seen.

\medskip

The upper plot of Fig.~\ref{fig:SSP_spectra} indicates that removing the troublesome SSP prior to the \texttt{STARLIGHT} fitting procedure makes essentially no discernible difference to the best-fit spectra of any of the three galaxies: for each of the three galaxies, the two best-fit spectra shown are almost identical to each other. Furthermore, the flux residuals for these galaxies tell a similar story: the residuals between the raw MaNGA spectrum and the two best-fit spectra for each galaxy are so similar as to be virtually indistinguishable.

In short, this test demonstrates that the best-fit spectrum produced by \texttt{STARLIGHT} for a given MaNGA galaxy is very unlikely to be affected by the removal of the lone SSP template at $\log(\rm age / yr) = 8.15$ prior to analysis.

\subsection{Comparison of Stellar Metallicities}
\label{subsec:testing_stellar_metallicities}

As explained in Section~\ref{subsec:Stellar}, the result of the \texttt{STARLIGHT} fitting process is a set of weights for the mass contributions of SSPs with different ages and chemical compositions to the light seen in the spectrum of each spaxel across the face of a galaxy. In this work, we use these weights to derive mass-weighted mean stellar metallicities and SFHs for the sample galaxies.

In order to test whether the mass-weighted mean stellar metallicities obtained in this work are reliable, we compared our values to those from two different Value Added Catalogues (VACs), both of which provide spatially-resolved stellar population properties for the full set of MaNGA galaxies included in the seventeenth (and final) SDSS-IV data release \citep[DR17;][]{Abdurrouf2022TheData}. One VAC uses the full spectral fitting code Fitting IteRativEly For Likelihood analYsis (\texttt{FIREFLY}; \citealp{Wilkinson2017FIREFLYCode}), and its data products are described in full by \citet{Goddard2017SDSS-IVType} and \citet{Neumann2022TheGalaxies}; the other VAC uses the software \texttt{Pipe3D} \citep{Sanchez2016Pipe3DFIT3D, Sanchez2016Pipe3DDataproducts}, and the data products are detailed fully in \citet{Lacerda2022PyFIT3DPipeline} and \citet{Sanchez2022SDSS-IVGalaxies}.

\medskip

\begin{figure*}
	\includegraphics[width=0.95\textwidth]{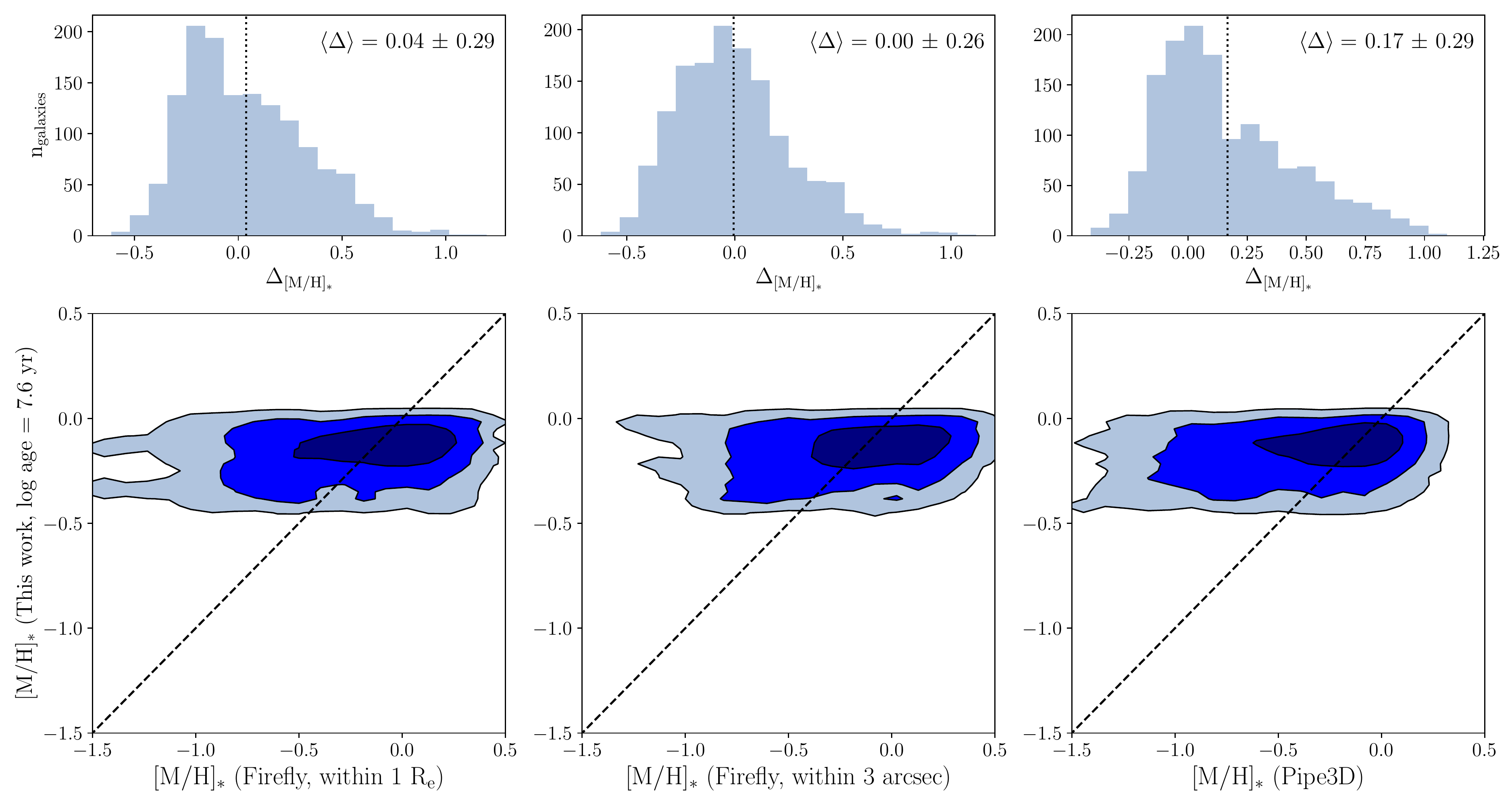}
    \caption{The vertical axis of these plots shows median mass-weighted stellar metallicities calculated for the sample galaxies in this work at a lookback time of $10^{7.6} \: \rm yr$. These data are compared with stellar metallicities calculated for the same galaxies provided by the \texttt{FIREFLY} VAC of DR17 galaxies \citep{Goddard2017SDSS-IVType, Neumann2022TheGalaxies} calculated within a shell located at 1 $\rm R_{e}$ (left-hand plots); with stellar metallicities provided by the same \texttt{FIREFLY} VAC calculated within a diameter of $3^{\prime\prime}$ (central plots); and finally with  stellar metallicities provided by the \texttt{Pipe3D} VAC of DR17 galaxies \citep[][right-hand plots]{Lacerda2022PyFIT3DPipeline, Sanchez2022SDSS-IVGalaxies}. The lower plots show confidence contours of $1 \sigma$, $2 \sigma$, and $3 \sigma$ for the data, with the 1-to-1 relation shown by the dashed black diagonal line. The corresponding upper plots show histograms plotting the density distribution of the difference between our median mass-weighted stellar metallicity data and those from the VACs cited above, $\Delta_{[\rm M / H]_{*}}$. The mean value of this difference, $\langle \Delta \rangle$, and its standard deviation, are also shown in these upper plots.}
    \label{fig:Stellar_metallicity_comparison_young}
\end{figure*}

\begin{figure*}
	\includegraphics[width=0.95\textwidth]{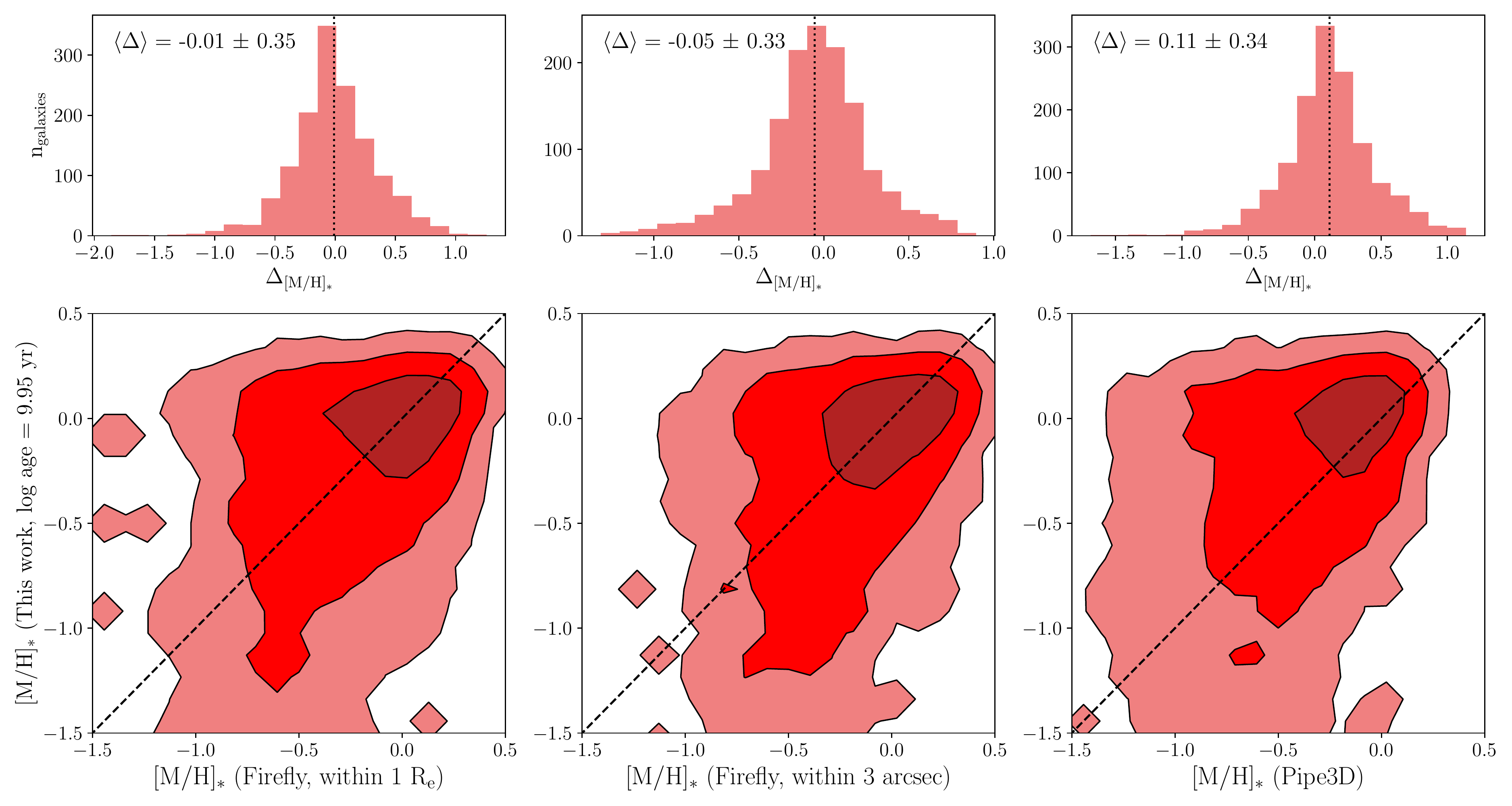}
    \caption{As Fig.~\ref{fig:Stellar_metallicity_comparison_young}, except that the median mass-weighted stellar metallicities plotted on the vertical axis are instead calculated for the sample galaxies in this work at a lookback time of $10^{9.95} \: \rm yr$. The data plotted on the horizontal axes are the same as in Fig.~\ref{fig:Stellar_metallicity_comparison_young}.}
    \label{fig:Stellar_metallicity_comparison_old}
\end{figure*}

Figures~\ref{fig:Stellar_metallicity_comparison_young}~and~\ref{fig:Stellar_metallicity_comparison_old} compare the median mass-weighted stellar metallicities calculated in this work with stellar metallicity values provided by the \texttt{FIREFLY} \citep{Goddard2017SDSS-IVType, Neumann2022TheGalaxies} and \texttt{Pipe3D} \citep{Lacerda2022PyFIT3DPipeline, Sanchez2022SDSS-IVGalaxies} VACs. The left-hand and central plots show stellar metallicity data for sample galaxies from the \texttt{FIREFLY} VAC within a shell located at 1 $\rm R_{e}$ and within a diameter of $3^{\prime\prime}$, respectively; the right-hand plots, meanwhile, show stellar metallicity data from the \texttt{Pipe3D} VAC. Confidence contours of $1 \sigma$, $2 \sigma$, and $3 \sigma$ are drawn for each of these comparative plots. The histograms plot the density distribution of the difference between the median mass-weighted stellar metallicities calculated in this work, and those from the various VAC data (denoted $\Delta_{[\rm M / H]_{*}}$). The mean value of this difference, $\langle \Delta \rangle$, is also quoted in each instance.

Both of these Figures display very similar data. Figure~\ref{fig:Stellar_metallicity_comparison_young} shows a comparison of the stellar metallicity data calculated in this work at a lookback time of $10^{7.6} \: \rm yr$ with the various VAC data; whereas Fig.~\ref{fig:Stellar_metallicity_comparison_old} instead plots a similar comparison of the stellar metallicity data calculated in this work at a lookback time of $10^{9.95} \: \rm yr$.

\medskip

At small ($10^{7.6} \: \rm yr$; Fig.~\ref{fig:Stellar_metallicity_comparison_young}) lookback times, corresponding to young stellar components, the stellar metallicity data obtained in this analysis are less well correlated with the various VAC data than at high ($10^{9.95} \: \rm yr$; Fig.~\ref{fig:Stellar_metallicity_comparison_old}) lookback times, corresponding to old stars. Indeed, at high lookback times, the correlation between our data and those of the various VACs improves significantly, lying close to the 1-to-1 relation and becoming a fairly tight relationship. This is not too surprising: at progressively younger stellar ages, the metallicities of a stellar population will become increasingly akin to the metallicity of the gas from which it is forming (cf. Section~\ref{subsec:MZR_comparison}), and the difficulties of deriving stellar metallicities for very young stars are equally difficult and uncertain for all the three methods compared here. Therefore, we would expect to see a better correlation between the stellar metallicity data in this work and those from the VACs at higher lookback times -- which is apparent in the different shapes of the confidence contours in Fig.~\ref{fig:Stellar_metallicity_comparison_young} and in Fig.~\ref{fig:Stellar_metallicity_comparison_old}.

Reassuringly, however, the mean value of $\Delta_{[\rm M / H]_{*}}$ is consistent with a mean of zero (within the range of the quoted uncertainty) for each of the six comparisons made across both sets of Figures at the different lookback times. This statistic implies that the mass-weighted mean stellar metallicities calculated during the \texttt{STARLIGHT} fitting process are in agreement with stellar metallicities provided by both the \texttt{FIREFLY} and \texttt{Pipe3D} VACs. The dispersion in the mean $\Delta_{[\rm M / H]_{*}}$ ranges from 0.26 to 0.35, which corresponds to ${\sim} 30 \%$ of the dynamical range covered by the sampled parameter. These mean $\Delta_{[\rm M / H]_{*}}$ statistics show higher dispersions than that found when comparing \texttt{FIREFLY} and \texttt{Pipe3D} data against each other; \citet{Sanchez2022SDSS-IVGalaxies} report a mean value of $\Delta_{[\rm M / H]_{*}} = 0.21 \pm 0.16$. The standard deviations reported are up to twice as high as those expected from simulations when using \texttt{STARLIGHT} (see both \citealp{CidFernandes2014} and \citealp{GonzalezDelgado2014InsightsSurvey}).

\subsection{Comparison of Star Formation Rates}
\label{subsec:testing_SFRs}

In this work, we use SFHs derived by the \texttt{STARLIGHT} fitting procedure to trivially determine the SFRs of the sample galaxies at different lookback times. It is therefore crucial to ensure that the SFRs used throughout this work are reasonably accurate. In order to test the reliability of these SFRs that we employ, we can produce a similar plot to those discussed in Section~\ref{subsec:testing_stellar_metallicities} and assess the data accordingly.

We can derive an alternative and independent value for the present-day SFR by measuring the intrinsic $\rm H \alpha$ flux -- provided by the MaNGA DAP -- for each of the sample galaxies. This procedure is described fully by \citet{Greener2020SDSS-IVGalaxies}, but in essence, we first correct the raw $\rm H \alpha$ flux for dust attenuation using measurements of the Balmer decrement (defined as the flux ratio of $\rm H \alpha$ to $\rm H \beta$) at each star-forming spaxel within a given galaxy and by employing the reddening curve of \citet{Calzetti2000TheGalaxies}. We then use the total galactic dust-corrected $\rm H \alpha$ flux in conjunction with luminosity distances provided by the NASA Sloan Atlas catalogue \citep{Blanton2011ImprovedImages} to obtain the $\rm H \alpha$ luminosity in Watts, $L_{{\rm H \alpha}} \: {\rm \left[ W \right]}$, for each of the sample galaxies. The $\rm H \alpha$ derived SFR, $\rm SFR_{H \alpha}$, for each galaxy is then calculated using the \citet{Kennicutt1998TheGalaxies} relation:
\begin{equation}
    {\rm SFR_{H \alpha}} \: \left[{\rm M_\odot \ yr^{-1}} \right] = \frac{L_{{\rm H \alpha}} \: {\rm \left[ W \right]}}{2.16 \times 10^{34} \: \rm M_\odot \ yr^{-1}}.
	\label{eq:SFR}
\end{equation}

Since we obtain an estimate for the $\rm H \alpha$ derived SFR via the measurement of nebular emission lines, it makes sense to compare $\rm SFR_{H \alpha}$ with the SFR calculated by \texttt{STARLIGHT} only at the youngest stellar age -- a lookback time of $10^{7.6} \: \rm yr$. This is not only because such emission lines are produced at the present day, but also because only the young type O and B stars will have temperatures sufficient to ionise the neutral hydrogen gas which surrounds them into \textsc{H~ii} regions, thus producing the nebular emission lines we require to measure the $\rm H \alpha$ derived SFRs of their host galaxies.

\medskip

\begin{figure}
	\includegraphics[width=0.45\textwidth]{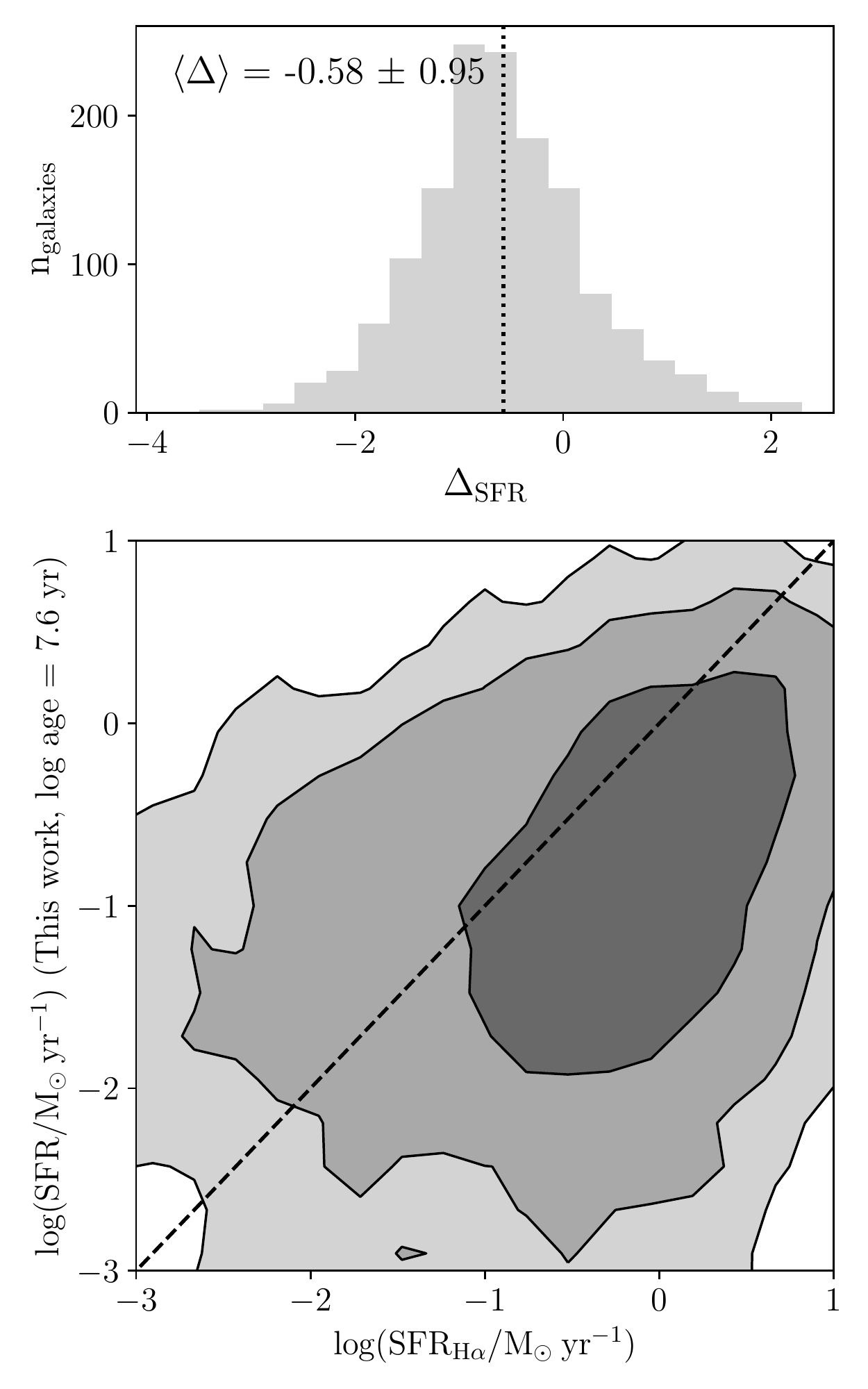}
    \caption{SFRs calculated for the sample galaxies in this work by using the \texttt{STARLIGHT} fitting procedure at a lookback time of $10^{7.6} \: \rm yr$ compared with SFRs derived from measurements of the $\rm H \alpha$ emission in these same galaxies. Similarly to Fig.~\ref{fig:Stellar_metallicity_comparison_young}, the lower plot shows confidence contours of $1 \sigma$, $2 \sigma$, and $3 \sigma$ for the data, with the 1-to-1 relation shown by the dashed black diagonal line. The upper plot shows a histogram plotting the density distribution of the difference between the two measures of SFR, $\Delta_{\rm SFR}$. The mean value of this difference, $\langle \Delta \rangle$, and its standard deviation, are also shown in this upper plot.}
    \label{fig:SFR_comparison}
\end{figure}

Figure~\ref{fig:SFR_comparison} therefore shows a comparison between these two SFR values for the sample galaxies, using the same layout as in Figs.~\ref{fig:Stellar_metallicity_comparison_young}~and~\ref{fig:Stellar_metallicity_comparison_old}. Confidence contours of $1 \sigma$, $2 \sigma$, and $3 \sigma$ are again displayed, and the histogram once again shows the density distribution of the difference between the two SFR values (which we denote $\Delta_{\rm SFR}$).

\medskip

The two independent SFR measurements are correlated, but the $\rm H \alpha$ derived SFRs are systematically slightly higher than the corresponding SFRs determined using \texttt{STARLIGHT}. Despite this offset, the 1-to-1 relation is found to pass within the inner confidence contour. The mean value of $\Delta_{\rm SFR}$ is equal to $-0.58 \pm 0.95$. The scatter in the mean value of this statistic is larger than that found in similar analyses undertaken by both \citet{Peterken2021SizeMaNGA} and \citet{Sanchez2022SDSS-IVGalaxies}, who report uncertainties of ${\sim} 0.7$ dex and 0.32 dex respectively, and is also larger than the ${\sim} 0.25$ dex standard deviation of the SFMS reported by \citet{Sanchez2022SDSS-IVGalaxies}.

Given the different systematic errors affecting these independent SFR estimates, a small offset is to be expected. This is a result also found by \citet[][see their Fig.~2]{Peterken2021SizeMaNGA}, who also compared SFRs derived by \texttt{STARLIGHT} with $\rm SFR_{H \alpha}$ and found the latter quantity to be systematically slightly higher than the former. Such an offset will not, however, affect any of the conclusions presented in this paper. As such, we may conclude that the SFRs used in this work are -- like the mass-weighted stellar metallicities produced by \texttt{STARLIGHT} -- also indeed reasonably reliable and safe to use.


\bsp	
\label{lastpage}
\end{document}